\documentclass[prb,a4paper,amsmath,amssymb,amsfonts,twocolumn]{revtex4}

\usepackage{epic,eepic}
\usepackage{graphicx}
\usepackage{subfigure}

\newcommand{\br}{\ensuremath{\mathbf{r}}}
\newcommand{\bs}{\ensuremath{\mathbf{s}}}
\newcommand{\bq}{\ensuremath{\mathbf{q}}}
\newcommand{\bk}{\ensuremath{\mathbf{k}}}
\newcommand{\nbk}{\ensuremath{n^{(\mathbf{k})}}}
\newcommand{\nsqbs}{\ensuremath{n^{(1,1)}(\mathbf{s})}}
\newcommand{\nhbs}{\ensuremath{n^{(1,0)}(\mathbf{s})}}
\newcommand{\nvbs}{\ensuremath{n^{(0,1)}(\mathbf{s})}}
\newcommand{\npbs}{\ensuremath{n^{(0,0)}(\mathbf{s})}}
\newcommand{\bFex}{\ensuremath{\beta \mathcal{F}_{{\rm ex}}}}
\newcommand{\Z}{\ensuremath{\mathbb{Z}}}
\newcommand{\ZZ}{\ensuremath{\mathbb{Z}^2}}
\newcommand{\ZZZ}{\ensuremath{\mathbb{Z}^3}}
\newcommand{\rhoa}{\ensuremath{\rho_\alpha}}

\newcommand{\dnsq}{
\setlength{\unitlength}{0.1em}
\begin{picture}(18,10)(-4,2)
\path(0,0)(10,0)
\path(10,0)(10,10)
\path(0,10)(10,10)
\path(0,0)(0,10)
\put(0,0){\whiten\circle{4}}
\put(10,0){\whiten\circle{4}}
\put(0,10){\whiten\circle{4}}
\put(10,10){\whiten\circle{4}}
\end{picture}}

\newcommand{\dnv}{
\setlength{\unitlength}{0.1em}
\begin{picture}(8,10)(-4,3)
\path(0,0)(0,10)
\put(0,0){\whiten\circle{4}}
\put(0,10){\whiten\circle{4}}
\end{picture}}

\newcommand{\dnh}{
\setlength{\unitlength}{0.1em}
\begin{picture}(18,6)(0,-2)
\path(4,0)(14,0)
\put(4,0){\whiten\circle{4}}
\put(14,0){\whiten\circle{4}}
\end{picture}}

\newcommand{\dnp}{
\setlength{\unitlength}{0.1em}
\begin{picture}(8,6)(0,-2)
\put(4,0){\whiten\circle{4}}
\end{picture}}

\newcommand{\dncb}{
\setlength{\unitlength}{0.1em}
\begin{picture}(22.5,10)(-4,4)
\path(0,0)(10,0)
\path(10,0)(10,10)
\path(0,10)(10,10)
\path(0,0)(0,10)
\path(0,10)(5.3,15.3)
\path(5.3,15.3)(15.3,15.3)
\path(15.3,15.3)(15.3,5.3)
\path(15.3,5.3)(10,0)
\path(15.3,15.3)(10,10)
\path(5.3,15.3)(5.3,5.3)
\path(5.3,5.3)(15.3,5.3)
\path(5.3,5.3)(0,0)
\put(0,0){\whiten\circle{4}}
\put(10,0){\whiten\circle{4}}
\put(0,10){\whiten\circle{4}}
\put(10,10){\whiten\circle{4}}
\put(5.3,15.3){\whiten\circle{4}}
\put(15.3,15.3){\whiten\circle{4}}
\put(15.3,5.3){\whiten\circle{4}}
\put(5.3,5.3){\whiten\circle{4}}
\end{picture}}

\newcommand{\dnxz}{
\setlength{\unitlength}{0.1em}
\begin{picture}(12.5,10)(-3,4)
\path(0,0)(0,10)
\path(0,10)(5.3,15.3)
\path(5.3,15.3)(5.3,5.3)
\path(5.3,5.3)(0,0)
\put(0,0){\whiten\circle{4}}
\put(0,10){\whiten\circle{4}}
\put(5.3,15.3){\whiten\circle{4}}
\put(5.3,5.3){\whiten\circle{4}}
\end{picture}}

\newcommand{\dnxy}{
\setlength{\unitlength}{0.1em}
\begin{picture}(22.5,6)(0,0)
\path(4,0)(14,0)
\path(4,0)(9.3,5.3)
\path(9.3,5.3)(19.3,5.3)
\path(19.3,5.3)(14,0)
\put(4,0){\whiten\circle{4}}
\put(14,0){\whiten\circle{4}}
\put(9.3,5.3){\whiten\circle{4}}
\put(19.3,5.3){\whiten\circle{4}}
\end{picture}}

\newcommand{\dnx}{
\setlength{\unitlength}{0.1em}
\begin{picture}(12.5,6)(0,0)
\path(4,0)(9.3,5.3)
\put(4,0){\whiten\circle{4}}
\put(9.3,5.3){\whiten\circle{4}}
\end{picture}}

\begin{document}
\title{Phase behavior of hard-core lattice gases: A Fundamental
Measure approach}
\date{\today}
\author{Luis Lafuente}
\email{llafuent@math.uc3m.es}

\author{Jos\'e A.\ Cuesta}
\email{cuesta@math.uc3m.es}

\affiliation{Grupo Interdisciplinar de Sistemas Complejos (GISC),
Departamento de Matem\'aticas,
Universidad Carlos III de Madrid, Avda.\ de la Universidad 30,
28911--Legan\'es, Madrid, Spain}

\begin{abstract}
We use an extension of fundamental measure theory to lattice hard-core fluids
to study the phase diagram of two different systems. First, two-dimensional
parallel hard squares with edge-length $\sigma=2$ in a simple square
lattice. This system is equivalent to the lattice gas with first and
second neighbor exclusion in the same lattice, and has the peculiarity
that its close packing is degenerated (the system orders in sliding columns).
A comparison with other theories is discussed. Second, a three-dimensional
binary mixture of parallel hard cubes with $\sigma_{\rm{L}}=6$ and
$\sigma_{\rm{S}}=2$. Previous simulations of this model only focused on
fluid phases. Thanks to the simplicity introduced by the discrete nature
of the lattice we have been able to map out the complete phase diagram
(both uniform and nonuniform phases) through a free minimization of the free
energy functional, so the structure of the ordered phases is obtained as
a result.  A zoo of entropy-driven phase transitions is found: one-, two-
and three-dimensional positional ordering, as well as fluid-ordered
phase and solid-solid demixings.
\end{abstract}

\maketitle
\section{Introduction}
Hard-core systems are the paradigm of entropy-driven
phase transitions. The first example of an entropy-driven (orientational)
ordering transition is given in the famous Onsager's paper\cite{onsager:1949} on
the isotropic-nematic transition in a three-dimensional system of thin hard
rods. But probably, the best known and discussed example of entropy-driven
(three-dimensional positional) ordering transition is the freezing of hard spheres.
This was first devised by Kirkwood et al.\cite{kirkwood:1950} using an approximate
theory, but the definite evidence about the existence of such a purely entropic
transition was the numerical simulations of Alder and Wainwright,\cite{alder:1957}
and Wood and Jacobson.\cite{wood:1957} As very few models can be solved exactly,
definite conclusions on the existence of phase transitions often come from numerical
simulations. But in many cases these are very demanding and powerful computers are
needed in order to reach a reliable system size. This fact, together with the
inexistence of appropriate theoretical approaches,
could explain that until the end of the eighties there were no more instances
of entropy-driven ordering transitions. At that time, a series of
numerical simulations\cite{frenkel:1988,veerman:1992,bolhuis:1997} 
showed that hard-core interaction can also induce one- and two-dimensional positional
ordering (smectic and columnar phases, respectively, in liquid crystal terminology).
This was a very striking fact, because it was generally believed
that the mechanism underlying these phase transitions was the decrease of internal energy
rather than the gain of entropy.

Apart from ordering transitions, it is well known that binary nonadditive mixtures
can demix by a pure entropic effect. An extreme case of nonadditivity was studied
by Widom and Rowlinson\cite{widom:1970} in a model with two different species
interacting ideally between members of the same species ($\sigma_{\rm{AA}}=\sigma_{\rm{BB}}=0$)
and with a hard-core interaction between unlike particles ($\sigma_{\rm{AB}}=\sigma$).
They rigorously showed that the system demixes into two fluid phases with different compositions. This can be
easily understood if we notice that the available volume is more effectively filled
by pure phases than by the mixture. Another interesting example of this kind is found
in colloid-polymer mixtures. Experimentally, it is well known that the addition of
non-adsorbing polymers to a colloidal suspension induces an effective attraction
between the colloidal particles that can induce the flocculation of the colloid.
A simple explanation for this effect is that the clustering of colloids (large particles)
leaves more free volume to the polymers (small particles), what translates into a
gain of entropy. This mechanism is known as {\em depletion}. Many models
\cite{asakura:1954,widom:1967,frenkel:1992} have been successfully introduced
in order to illustrate how
this effect can induce a fluid-fluid phase separation in mixtures.

Special mention merits the case of the additive binary mixture of hard spheres.
The absence of a spinodal instability in the Percus-Yevick solution for this system\cite{lebowitz:1964}
led to believe that entropic demixing was not possible for additive mixtures.
But almost thirty years later, Biben and Hasen\cite{biben:1991} predicted such a spinodal
by using a more accurate integral equation theory.
Since this result, a bunch of theoretical,\cite{lekkerkerker:1993,
rosenfeld:1994,poon:1994,caccamo:1997,coussaert:1998,velasco:1999}
simulation\cite{buhot:1998,dijkstra:1998a,dijkstra:1999,garcia-almarza:1999} and
experimental\cite{kaplan:1994,dinsmore:1995,steiner:1995,imhof:1995} results appeared
supporting the existence of demixing in additive
binary mixtures of hard spheres when the diameter ratio is at least $5:1$. Almost at the same time,
it was pointed out that instead of a fluid-fluid demixing at least one of the separated phases might be
ordered.\cite{poon:1994,caccamo:1997,coussaert:1998,velasco:1999,
dijkstra:1998a,dijkstra:1999,garcia-almarza:1999,dinsmore:1995,steiner:1995,imhof:1995}
The actual scenario for this system is a metastable fluid-fluid demixing (not confirmed
by direct simulation\cite{dijkstra:1999}) preempted by a fluid-solid coexistence or (if the mixture is sufficiently
asymmetric) a solid-solid one. Qualitatively, this is the same situation one finds in a binary additive
mixture of parallel hard cubes.\cite{cuesta:1996,martinez-raton:1998,martinez-raton:1999}

{}From a theoretical point of view one
of the first exactly solvable hard-core models showing a fluid-solid transition
was a lattice model proposed by Temperley.\cite{temperley:1965} Many other lattice
hard-core models were studied in the sixties by adapting the approximate
theories developed for Ising-like models to hard body systems.\cite{burley:1972,runnels:1972}
They succeeded in the prediction of an order-disorder transition and mainly focused
on studying the dependence of the nature of the transition upon the range of the hard-core and the
topology of the underlying lattice.

For the continuum model of hard spheres, one of the most
successful theories to study the freezing has been density functional theory (DFT). Many accurate
functionals have been devised for the monocomponent fluid,\cite{evans:1992} but when
they are applied to the binary mixture some problems arise: (i) many of the theories employed are not
directly formulated for mixtures and the extension is far from being straightforward,
(ii) it is very difficult to study the solid phases because it is not trivial to
determine which is the most stable structure for the mixture, and this information is an input
in most approaches.\cite{denton:1990} These difficulties have been circumvented by mapping the
binary hard-core mixture into a monocomponent fluid (large particles alone) with a hard-core
and an effective short-range attractive potential. It is then possible to use perturbation
theory in order to study the phase diagram.\cite{velasco:1999} On the other hand, the solid phase is usually
assumed to be an fcc crystal of the large particles with the small particles
uniformly distributed. Although this approach has been extensively
used,\cite{dijkstra:1998a,dijkstra:1998b,garcia-almarza:1999,louis:2000,germain:2002}
it is only valid for low molar fractions of the small particles. Besides, even in this
case, the assumption that the density of small particles is uniform in the ordered phase is
rather unrealistic because the ordering of large particles induces structure in the distribution of
the small ones. To the best of our knowledge, this problem has not been addressed satisfactorily yet.

A direct study (without resorting to an effective one-component fluid) has been carried out
for a binary mixture of parallel hard cubes with Rosenfeld's fundamental measure
theory\cite{cuesta:1996,martinez-raton:1998,martinez-raton:1999} (FMT). This theory
has the advantage of being naturally formulated for mixtures. A complete analysis
of fluid-fluid demixing has been performed for arbitrary size ratios, but again, the
lack of intuition about the distribution of small particles in the crystal makes impossible
to study freezing in this system. To solve this problem one should perform a free minimization
of the free energy functional and obtain the structure of the ordered phases as an output.
But due to the continuum nature of the system this would require a huge amount of numerical
work.

The situation is more favorable for the lattice counterpart of this model. Indeed,
simulations of a binary mixture of parallel hard cubes ($6:2$) on a
simple cubic lattice were performed by Dijkstra and Frenkel,\cite{dijkstra:1994} but
their focus was whether entropic demixing could be observed in additive binary mixtures
and the structure of the inhomogeneous phases was not considered. The results of these simulations
(a stable fluid-fluid demixing) are in contradiction with the predictions
of the continuum system (it exhibits a fluid-fluid spinodal only for size ratios above $10:1$,
and it is always preempted by freezing of the large component). With the aim of explaining
this mismatch, we extended the fundamental measure functional for parallel hard cubes to
the lattice version.\cite{lafuente:2002b} With this theory we have shown in a previous
work\cite{lafuente:2002a} that the latter is the correct picture. (A more detailed account of
this work will be given here.) Furthermore, due to the discrete nature of the system it
is possible to give a complete description of the ordered phases (see below).
Thus, we show that lattice models,
treated in a suitable manner, can serve as a starting point to study the structure of ordered
phases in (continuum) mixtures.

There is a second benefit of this extension of FMT to lattices that we want to emphasize.
These simulations, together with an exactly solvable model
proposed by Widom\cite{widom:1967} and Frenkel and Louis\cite{frenkel:1992} in different contexts,
show that lattice models can give accurate descriptions of demixing phenomena. But in
spite of their historical role in the development of Statistical Mechanics and their simplicity,
with a few exceptions,\cite{nieswand:1993a,nieswand:1993b,reinel:1994} density functional theories
have only focused on continuum models. We believe that the formulation of classical density
functional approaches for lattice models will help to better understand both, the phase
behavior of complex fluids and the formal structure of the approximate functionals.

The paper is organized as follows.
A review of the lattice version of FMT is presented in Sec.~\ref{LFMT}. In Sec.~\ref{app},
we use this theory to obtain the complete phase diagram of
two systems. First, a two-dimensional system of parallel hard squares 
with edge-length $\sigma=2$ on a square lattice
(this is equivalent to the two-dimensional lattice gas with
first and second neighbor exclusion). This system has been widely studied in the literature
(see Refs.~\onlinecite{runnels:1972} and \onlinecite{burley:1972} and references therein)
and there exists a big controversy about its phase behavior
so far unsettled. A detailed analysis is performed by applying
the new theory, and a comparison with results from
other theories is discussed. The lattice fundamental
measure theory (LFMT) appears to be at the same level of accuracy of the other well accepted theories.
Secondly, we have addressed the problem of the binary additive mixture studied by simulations, i.e.\
a binary mixture of parallel hard cubes ($\sigma_{\rm{L}}=2$, $\sigma_{\rm{S}}=6$) on a simple cubic
lattice. Due to the simplification introduced by the lattice, the complete phase diagram has been
mapped out. It shows a very rich collection of entropic phase transitions. 
As a matter of fact, we have found one-, two- and
three-dimensional ordering transitions, as well as
fluid-ordered phase and solid-solid demixings.
A free minimization of the free energy functional has been performed, so the structure of the ordered
phases has also been obtained. Finally, conclusions are discussed in Sec.~\ref{conclusions}.

\section{Theory}\label{LFMT}

The construction of the lattice fundamental measure 
functional is based on the ideas of
the exact zero-dimensional reduction \cite{tarazona:1997,cuesta:1997a} and in
the exact form of the one-dimensional functional. A full
account of the details of the procedure can be found in
Ref.~\onlinecite{lafuente:2002b}. In that work, the general form of the functional
for a system of parallel hard cubes in a simple cubic lattice
(for any dimension, particle size or number of components)
is presented. Basically, the idea behind it
is to construct a family of functionals for arbitrary dimension
in such a way that they consistently satisfy the dimensional reduction
property of the exact functionals down to zero-dimensional
cavities (i.e.\ cavities which can host no more than one
particle). Moreover, the prescription chosen is inspired in
the exact functional for the one-dimensional system, which is
recovered with the scheme proposed.

Let us consider a $d$-dimensional additive mixture of parallel hypercubes with edge-lengths
$\sigma_{\alpha}=2a_{\alpha}+\epsilon$ lattice spacings, where $\alpha$
is the species index and $\epsilon=0,1$ does not depend on $\alpha$, i.e all the
species have, simultaneously, even or odd sizes (the mixed---nonadditive---case
is more involved,\cite{lafuente:2002b} so we just ignored it because 
it will not be used anywhere in this work). In
Ref.~\onlinecite{lafuente:2002b} the excess free energy functional
for this system in this approximation
was found to be [cf.\ Eq.~(3.2) of that reference]
\begin{equation}\label{functional}
\bFex[\rho]=\sum_{\bs \in \Z^d}\sum_{\bk \in \{0,1\}^d} (-1)^{d-k}
\Phi_0\left(\nbk(\bs)\right),
\end{equation}
where $k=\sum_{l=1}^d k_l$, $\Phi_0(\eta)=\eta+(1-\eta) \ln (1-\eta)$
is the excess free energy for a zero-dimensional cavity with mean
occupancy $0\eta\leq1$,
$\beta$ the reciprocal temperature in Boltzman's units
and $\nbk(\bs)$ are weighted densities
defined by the convolutions
\begin{equation}\label{wd}
\nbk(\bs)=\sum_{\alpha}\sum_{\br\in\Z^d}w_{\alpha}^{(\bk)}(\bs-\br)\rhoa(\br),
\end{equation}
$\rhoa(\bs)$ being the one-particle distribution function for
species $\alpha$ and
\begin{eqnarray}
w_{\alpha}^{(\bk)}(\bs)&=&\prod_{l=1}^d w_{\alpha}^{(k_l)}(s_l),\\
w_{\alpha}^{(k)}(s)&=&\left\{
\begin{array}{ll}
1&\text{if $-a_{\alpha}-k-\epsilon<s<a_{\alpha}$,}\\
0&\text{otherwise.}
\end{array}
\right.
\end{eqnarray}
Notice that as weights are indexed by $\bk\in\{0,1\}^d$, there are
$2^d$ different weighted densities.

The direct correlation function between species $\alpha$ and $\gamma$ can
be obtained from this functional as
\begin{equation*}
c_{\alpha\gamma}(\bs-\br)=
-\frac{\partial^2\bFex[\rho]}{\partial\rho_{\alpha}(\bs)\partial\rho_{\gamma}(\br)}
\bigg|_{\text{uniform}}.
\end{equation*}
Then, from (\ref{functional}),
\begin{equation}\label{dcf}
c_{\alpha\gamma}(\bs)=
-\sum_{\bk\in\{0,1\}^d}\frac{(-1)^{d-k}}{1-n_k}
\varphi_{\alpha\gamma}^{(\bk)}(\bs),
\end{equation}
where $n_k=\sum_{\alpha}\sigma_{\alpha}^k(\sigma_\alpha-1)^{d-k}\rho_{\alpha}$
are the weighted densities (\ref{wd}) in the uniform limit and
$\varphi_{\alpha\gamma}^{(\bk)}(\bs)$ is the convolution
\begin{equation}
\varphi_{\alpha\gamma}^{(\bk)}(\bs)\equiv\sum_{\br\in\Z^d}w_{\alpha}^{(\bk)}(\br)
w_{\gamma}^{(\bk)}(\br+\bs).
\end{equation}
Because of the structure of the direct correlation function it is
convenient to work with its discrete Fourier transform, which takes
the form
\begin{equation}
\hat{c}_{\alpha\gamma}(\bq)=-\sum_{\bk\in\{0,1\}^d}\frac{(-1)^{d-k}}{1-n_k}
\hat{w}^{(\bk)}_{\alpha}(-\bq)\hat{w}^{(\bk)}_{\gamma}(\bq),
\end{equation} 
where
\begin{equation}
\hat{w}^{(\bk)}_{\alpha}(\bq)=\prod_{l=1}^{d}{\rm e}^{-i \frac{q_l}{2} k_l}\frac{\sin q_l
(a_{\alpha}+\frac{k_l-1}{2})}{\sin q_l/2}.
\end{equation}

The general expression of the functional (\ref{functional}) adopts
very simple forms when particularized to specific systems. In
order to make clear the structure of the functional, we will
introduce a diagrammatic notation which helps visualizing
its dimensional reduction properties in a simple
way. For the sake of simplicity let us consider the lattice gas with first
and second neighbor exclusion in a two-dimensional square lattice.
This is a system of parallel hard squares with $\sigma=2$ lattice spacings.
In diagrammatic notation the excess free energy functional (\ref{functional})
can be written
\begin{equation}\label{f2d}
\bFex[\rho]=\sum_{\bs \in \ZZ} [\Phi_0\left(\dnsq\right)
-\Phi_0\left(\dnh\right)-\Phi_0\left(\dnv\right)+\Phi_0\left(\dnp\right)],
\end{equation}
where the diagrams represent
the weighted densities (\ref{wd}) as
\begin{eqnarray}\label{diag2d}
\dnsq=\nsqbs&=&\rho(s_1,s_2)+\rho(s_1+1,s_2)\nonumber \\
&&+\rho(s_1,s_2+1)+\rho(s_1+1,s_2+1),\nonumber \\
\dnh=\nhbs&=&\rho(s_1,s_2)+\rho(s_1+1,s_2), \\
\dnv=\nvbs&=&\rho(s_1,s_2)+\rho(s_1,s_2+1),\nonumber \\
\dnp=\npbs&=&\rho(s_1,s_2).\nonumber
\end{eqnarray}

What becomes apparent with this diagrammatic notation is that the excess
functional (\ref{functional}) can be regarded as a linear combination of contributions
due to a particular set of zero-dimensional cavities (\ref{diag2d}). Furthermore,
we can manipulate the diagrams in order to prove the dimensional reduction properties
that the functional (\ref{f2d}) satisfies. To illustrate this, we will consider the
dimensional reduction to a one-dimensional system, the hard rod lattice gas, whose
exact excess functional is known to have the form
(\ref{functional}).\cite{lafuente:2002b} To perform this reduction,
we will apply an infinite external potential in every site of the lattice except
in an infinite linear chain defined by $\mathcal{L}=\{(s_1,0):s_1\in\Z\}$. This implies that
the centers of mass of the particles can only occupy the sites in $\mathcal{L}$, the
system becoming equivalent to a hard rod lattice gas with particles of 
size $\sigma=2$.
In terms of $\rho(\bs)$ this means that $\rho(\bs)=\rho(s_1)\delta_{s_2,0}$,
where $\delta_{i,j}$ is the Kronecker symbol and $\rho(s)$ is the one-particle
distribution function for the one-dimensional system. Within this constraint,
the excess free-energy functional of the effective system can be obtained by summing over
$s_2\in\Z$ in (\ref{f2d}). Each contribution gives, respectively,
\begin{eqnarray*}
\sum_{s_2\in\Z}\Phi_0\left(\dnsq\right)&=&2\Phi_0\left(\dnh\right),\\
\sum_{s_2\in\Z}\Phi_0\left(\dnh\right)&=&\Phi_0\left(\dnh\right),\\
\sum_{s_2\in\Z}\Phi_0\left(\dnv\right)&=&2\Phi_0\left(\dnp\right),\\
\sum_{s_2\in\Z}\Phi_0\left(\dnp\right)&=&\Phi_0\left(\dnp\right),
\end{eqnarray*}
where the diagrams in the r.h.s.\ must now be interpreted as
$\dnh=\rho(s)+\rho(s+1)$ and $\dnp=\rho(s)$. Therefore,
the excess free-energy functional
for the one-dimensional system so obtained is
\begin{equation}
\bFex^{(1d)}[\rho]=\sum_{s_1\in\Z}[\Phi_0\left(\dnh\right)
-\Phi_0\left(\dnp\right)],
\end{equation}
which coincides with the exact result [see Eqs.~(2.26) and (3.1) of
Ref.~\onlinecite{lafuente:2002b}].

Another example in three-dimensions is given in the Appendix.
\section{Applications}\label{app}

In spite of the simple structure of the lattice fundamental measure
functionals, the applications to specific systems have proven able
to describe very complex phase diagrams.\cite{lafuente:2002a}
In this section we will study in detail two particular systems:
first, the lattice gas with exclusion to first and second neighbors
on a square lattice; second, a binary mixture
of parallel hard cubes with $\sigma=2$ and $6$,
in a simple cubic lattice.

\subsection{Parallel hard squares}

This model is defined by the interaction pair potential 
\begin{equation}
\phi(\bs,\bs')=\left\{
\begin{array}{ll}
\infty&\text{if $|s_i-s'_i|\leq1$\, for both $i=1,2$,}\\
0&\text{otherwise.}
\end{array}\right.
\end{equation}
It has been previously studied employing other approximate theories,
such as finite-size scaling methods,\cite{ree:1967,nisbet:1974} series expansions
\cite{bellemans:1967,baram:1983,baram:1987} and clusters methods.
\cite{bellemans:1967,burley:1972}
All authors agree in that the close-packed state
is a columnar phase (ordered along one dimension but
fluid along the other). This notwithstanding, the nature of the transition
remains doubtful, the results being highly dependent on the theory used.
\cite{runnels:1972}
While some authors conclude that the system exhibits a third order
transition very near close packing,\cite{bellemans:1967,ree:1967,baram:1983,baram:1987}
others obtain a second order transition at a lower density,
\cite{bellemans:1967,burley:1972,slotte:1983} and even
some of them have speculated about the lack of such a phase change.
\cite{bellemans:1967,nisbet:1974,slotte:1983}
The results obtained with the present theory are in accordance
with those of the second order phase transition. Unfortunately,
we have no concluding arguments to umpire this dispute.

The advantage of our approach over other theories is that it
provides a simple prescription to build a density functional
in closed form.
Then, all the powerful tools of density functional theory may
be applied. The excess functional for this system within the
present theory is that of Eq.~(\ref{f2d}).

In order to pin down the phase diagram of the system we will
proceed systematically: first, studying the uniform phases,
and second, considering spatial inhomogeneities.
For a uniform density profile $\rho(\bs)=\rho$, the weighted
densities (\ref{diag2d}) become $\dnsq=4\rho,
\dnh=\dnv=2\rho$ and
$\dnp=\rho$, where $0\leq\rho\leq1/4$.
The excess free energy density (in $k_{{\rm B}}T$ units) can be
calculated particularizing (\ref{f2d}), which yields
\begin{equation}
\Phi_{{\rm ex}}(\rho)=\Phi_0(4\rho)-2\Phi_0(2\rho)+\Phi_0(\rho),
\end{equation}
Adding up the ideal gas contribution, $\Phi_{{\rm id}}=\rho(\ln\rho-1)$,
and taking into account the definition of $\Phi_0$, we obtain the following
free energy density for the fluid
\begin{equation}\label{fed2d}
\begin{split}
\Phi_{{\rm fluid}}(\rho) =&\,\rho\ln \rho+(1-\rho)\ln(1-\rho)\\
&+(1-4\rho)\ln(1-4\rho) \\
&-2(1-2\rho)\ln(1-2\rho).
\end{split}
\end{equation}
{}From this, all the thermodynamic properties of
the fluid phase can be derived. For instance, the fugacity is given by
\begin{equation}
z_{{\rm fluid}}=\frac{\rho(1-2\rho)^4}{(1-\rho)(1-4\rho)^4},
\end{equation}
and the pressure takes the simple form
\begin{equation}\label{eos2d}
\beta p_{{\rm fluid}}=\ln\left[\frac{(1-2\rho)^2}{(1-\rho)(1-4\rho)}
\right].
\end{equation}

The structure of the equilibrium phase can be analyzed by means of
the direct correlation function, obtained from (\ref{dcf})
particularizing for $d=2$ and a single
component with $\sigma=2$. In a symmetry broken continuous
phase transition, for some $\bq\neq\mathbf{0}$
\begin{equation}\label{sf2d}
1-\rho \hat{c}(\bq)=0,
\end{equation}
this condition being equivalent to the divergence of the structure
factor. Since we are interested in the spatial instabilities of
the uniform phase, we have to look for the lowest value of $\rho$
which makes  the condition (\ref{sf2d}) solvable for some $\bq$.
Taking into account that the symmetry
of the system enables us to take $\bq=(q,0)$, Eq.~(\ref{sf2d})
becomes
\begin{equation}
\cos^2 (q/2)=-\frac{1-\frac{4\rho}{1-2\rho}+\frac{\rho}{1-\rho}}
{4\rho\left(\frac{4}{1-4\rho}-\frac{1}{1-2\rho}\right)}.
\end{equation}
Since the denominator is positive in the whole range of $\rho$, and so is the
numerator for small values of the density, the solution corresponds to
the vanishing of the latter. This occurs at the density
\begin{equation}
\rho_{{\rm crit}}=\frac{3-\sqrt{5}}{4}\approx0.1910,
\end{equation}
and, of course, $q=\pi$, implying the periodicity of the inhomogeneous
phase to be $d=2\pi/q=2$ lattice spacings.

\begin{figure}
\includegraphics[width=60mm,clip=]{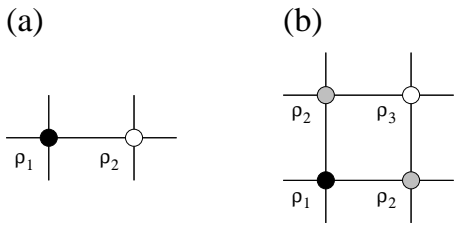}
\caption{\label{fig1}The unit cell for a columnar phase with
periodicity equal to two lattice spacings is shown in (a), and
for a solid phase with the same periodicity in (b).}
\end{figure}

What remains to be determined is the symmetry of the nonuniform phase at the
transition point. Based on the previous results for this system
and on those recently obtained for parallel hard cubes in
the continuum,\cite{groh:2001} we guess that this phase must
be either a columnar or a solid (ordering along the two coordinate
axes). In order to determine which phase is the stable one, we have 
performed a global minimization of the functional (\ref{f2d})
within the constraints imposed by the symmetry and periodicity
of both the columnar and the solid phases. For a generic columnar phase
with periodicity equal to two lattice spacings, the one-particle distribution
function takes the form
\begin{equation}\label{col2d}
\rho_{{\rm col}}(\bs)=\left\{
\begin{array}{ll}
\rho_1&\text{if $s_1$ is even},\\
\rho_2&\text{otherwise},
\end{array}
\right.
\end{equation}
while for a solid phase with the same periodicity we have
\begin{equation}\label{sol2d}
\rho_{{\rm sol}}(\bs)=\left\{
\begin{array}{ll}
\rho_1&\text{if $s_1$ and $s_2$ are even},\\
\rho_2&\text{if $s_1$ or $s_2$ is odd},\\
\rho_3&\text{otherwise}.
\end{array}
\right.
\end{equation}
A sketch of the unit cell for each case is shown in Fig.~\ref{fig1}.
Note that the uniform phase is included in both (\ref{col2d})
and (\ref{sol2d}).

For the columnar phase the total free energy density takes the form
\begin{equation}
\begin{split}
\Phi_{{\rm col}}(\rho_1;\rho)=&\,\Phi_{{\rm id}}(\rho_1,2\rho-\rho_1)+
\Phi_0(4\rho)-\Phi_0(2\rho)\\
&-\frac{1}{2}[\Phi_0(2\rho_1)+\Phi_0(4\rho-2\rho_1) \\
&-\Phi_0(\rho_1)-\Phi_0(2\rho-\rho_1)],
\end{split}
\end{equation}
where we have substituted the density profile (\ref{col2d}) in
(\ref{f2d}), used the relation $2 \rho=\rho_1+\rho_2$,
and introduced the ideal term $\Phi_{{\rm id}}(\rho_1,\rho_2)=\frac{1}{2}
\sum_i\rho_i(\ln\rho_i-1)$.

We can now minimize the total free energy density
at constant $\rho$. Note that in this case we have to minimize
with respect to a single variable.
The Euler-Lagrange equation is
\begin{equation}\label{el2dcol}
\frac{\rho_1 (1-2\rho_1)^2(1-2\rho+\rho_1)}
{(2\rho-\rho_1)(1-4\rho+2\rho_1)^2(1-\rho_1)}=1.
\end{equation}
One solution corresponds to the uniform phase ($\rho^{{\rm eq}}_1=\rho$). It
can be easily checked that this is indeed the minimum of the
free energy for $\rho<\rho_{{\rm crit}}$, as expected. After
removing this solution from (\ref{el2dcol}), we obtain a quadratic polynomial
whose roots become physical for $\rho\geq\rho_{{\rm crit}}$. Above the critical
density the uniform phase is no longer a minimum; instead, we have
a columnar phase given by
\begin{equation}
\rho_1^{{\rm eq}}=\rho+\frac{1}{2}
\sqrt{\frac{(1-2\rho)(\rho-\rho_{{\rm crit}})
(3-2\rho-2\rho_{{\rm crit}})}{\rho}}
\end{equation}
(we have chosen $\rho_1^{{\rm eq}}>\rho_2^{{\rm eq}}
=2\rho-\rho_1^{{\rm eq}}$).
This phase has a lower free energy than the fluid phase for
$\rho>\rho_{{\rm crit}}$, but we still have to calculate free energy for the
solid branch in order to know which one is the stable phase above the transition point.
\begin{figure}
\includegraphics[width=70mm,clip=]{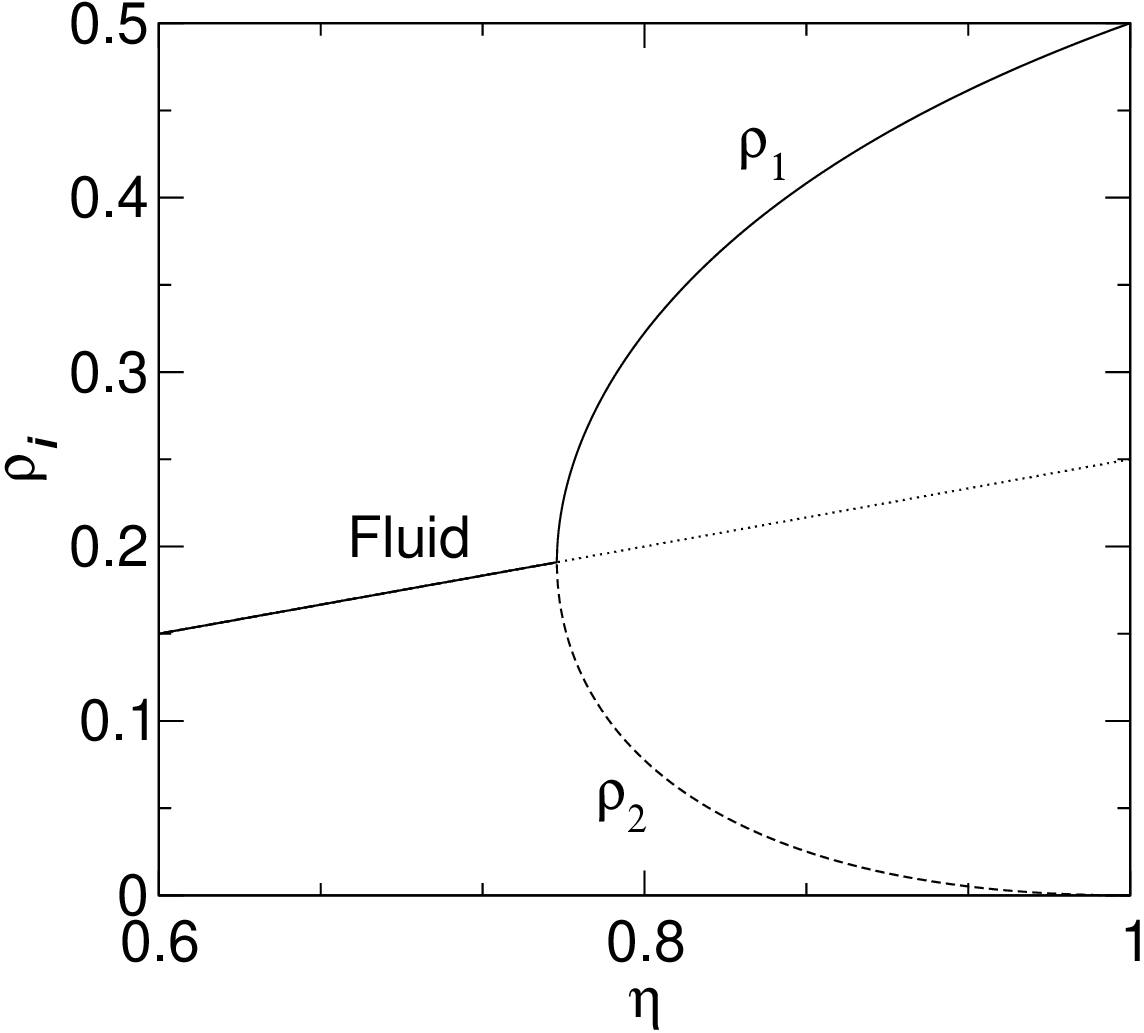}
\caption{\label{fig2} Sublattice densities for the columnar phase. It is also
shown the metastable fluid beyond the transition point.}
\end{figure}

\begin{figure}
\includegraphics[width=70mm,clip=]{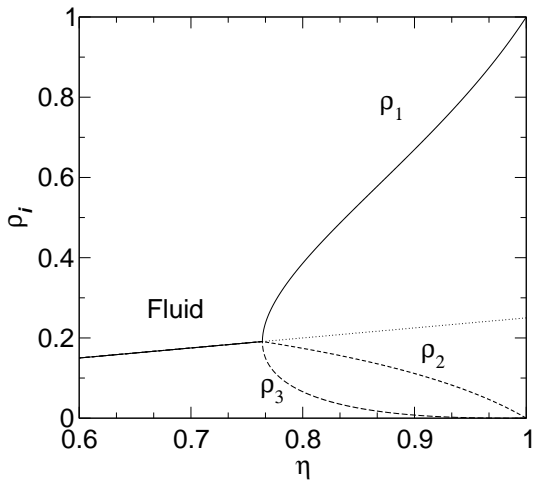}
\caption{\label{fig3} Sublattice densities for the solid phase. It is also
shown the metastable fluid beyond the transition point.}
\end{figure}

\begin{figure}
\includegraphics[width=70mm,clip=]{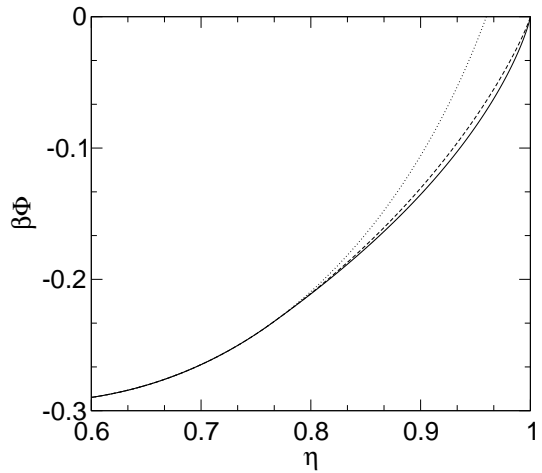}
\caption{\label{fig4} Free energy density of the fluid (dotted line), columnar (solid line)
and solid (dashed line) phases.}
\end{figure}

For the solid phase, substituting (\ref{sol2d}) in (\ref{f2d}) and adding the ideal
contribution, the total free energy density turns out to be
\begin{equation}\label{fed2dsol}
\begin{split}
\Phi_{{\rm sol}}(\rho_1,\rho_3;\rho)=&\,\Phi_{{\rm id}}(\rho_1,
2\rho-\rho_{+},\rho_3)+\Phi_0(4\rho) \\
&-\Phi_0(2\rho+\rho_{-})-\Phi_0(2\rho-\rho_{-})\\
&+\frac{1}{4}\left[\Phi_0(\rho_1)+2\Phi_0(2\rho-\rho_{+})+
\Phi_0(\rho_3)\right],
\end{split}
\end{equation}
where we have used $4\rho=\rho_1+2\rho_2+\rho_3$ to eliminate the dependence on
$\rho_2$, and have defined $\rho_{\pm}=(\rho_1\pm\rho_3)/2$. As in the previous
case, the equilibrium density profile is the
global minimum of (\ref{fed2dsol}) at constant $\rho$, but now we have two
independent variables, $\rho_1$ and $\rho_3$. Thus, the Euler-Lagrange
equations are now the system of algebraic equations
\begin{equation}
\begin{split}
\frac{\rho_1(1-2\rho-\rho_{-})^2(1-2\rho+\rho_{+})}{
(1-\rho_1)(2\rho-\rho_{+})(1-2\rho+\rho_{-})^2}&=1,\\
\frac{\rho_3(1-2\rho+\rho_{-})^2(1-2\rho+\rho_{+})}{
(1-\rho_3)(2\rho-\rho_{+})(1-2\rho-\rho_{-})^2}&=1.
\end{split}
\end{equation}
The fluid phase, given by $\rho_1^{{\rm eq}}=\rho_3^{{\rm eq}}=\rho$, is the
solution for $\rho\leq\rho_{{\rm crit}}$. The solution for $\rho\geq\rho_{{\rm crit}}$
must be obtained numerically and is plotted in Fig.~\ref{fig3}. In Fig.~\ref{fig4}
we can see that the solid branch bifurcates with a free energy lower than that of the
fluid phase, but larger than that of the columnar phase. The
transition is then fluid-columnar.

It is feasible to study analytically the behavior of each branch at the transition point. This
would give a definite conclusion about the nature of the phase change.
It is straightforward to check the continuity of ${{\rm d}}\Phi/{{\rm d}}\rho$ at $\rho_{\rm{crit}}$
(for both the solid and the columnar branches),
but a discontinuity is found in the second derivative at $\rho_{\rm{crit}}$, so the transition
is second order. Furthermore, the stable phase beyond the transition point is the
one with lowest second derivative for $\rho\to\rho_{\rm{crit}}^{+}$. From
the values
\begin{eqnarray}
\Phi_{{\rm fluid}}''(\rho_{\rm{crit}}^{+})&=&2(15+7 \sqrt{5})\approx61.3, \nonumber \\
\Phi_{{\rm col}}''(\rho_{\rm{crit}}^{+})&=&2(5+2 \sqrt{5})\approx18.9,\\
\Phi_{{\rm sol}}''(\rho_{\rm{crit}}^{+})&=&4(5+\sqrt{5})\approx28.9, \nonumber
\end{eqnarray}
we conclude that indeed the system undergoes a second order transition
from a fluid phase to a columnar one at $\rho_{{\rm crit}}$. Besides, as it
can be inferred from the density dependence of the free energy density
for every branch (Fig.~\ref{fig4}), the columnar phase remains the most
stable phase up to close packing.

\begin{figure}
\includegraphics[width=70mm,clip=]{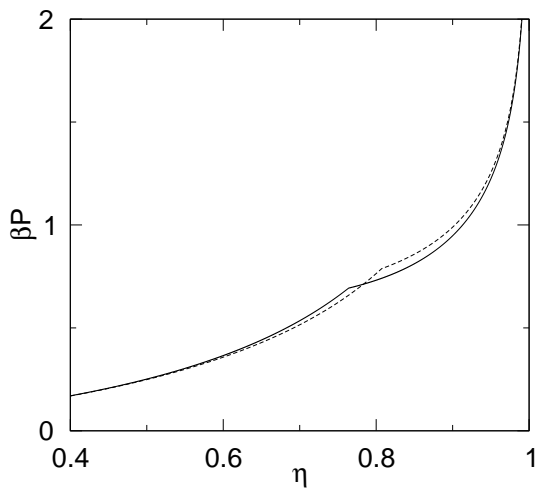}
\caption{\label{fig5} Equation of state of the hard square fluid from
lattice fundamental measure theory (solid line) and from the cluster method 
of Rushbrooke and Scoins (dashed line).}
\end{figure}

We can now plot the equation of state (Fig.~\ref{fig5}),
with the fluid branch given by Eq.~(\ref{eos2d}) and the columnar one
by
\begin{equation}\label{p2dcol}
\beta p_{{\rm col}}=\frac{1}{2} \ln \left[
\frac{(1-2\rho)^2 (1-2\rho_1^{{\rm eq}}) (1-4\rho+2\rho_1^{{\rm eq}})}
{(1-4\rho)^2 (1-\rho_1^{{\rm eq}}) (1-2\rho+\rho_1^{{\rm eq}})}\right].
\end{equation}
The fugacity of the columnar phase is given by
\begin{equation}\label{z2dcol}
z_{{\rm col}}=\frac{\rho_1^{{\rm eq}}(1-2\rho)^2(1-2\rho_1^{{\rm eq}})^2}
{(1-4\rho)^4(1-\rho_1^{{\rm eq}})}.
\end{equation}
At the critical point, we have
\begin{equation}
\beta p_{{\rm crit}}=\ln 2,\quad z_{{\rm crit}}=\frac{11+5\sqrt{5}}{2}.
\end{equation}

As mentioned at the beginning of this section, the results from this lattice fundamental
measure theory are compatible with the ones obtained by Bellemans and Nigam\cite{bellemans:1967}
($\rho_{{\rm crit}}\approx0.202$, $\beta p_{{\rm crit}}\approx 0.788$
and $z_{{\rm crit}}\approx17.29$) through the cluster method
of Rushbrooke and Scoins\cite{rushbrooke:1955} (plotted with a dashed line
in Fig.~\ref{fig5}). From Fig.~\ref{fig5} we can see that the agreement at low and high
densities is very accurate, and deviations occur only near the critical point.
This can be understood if we realize that both theories neglect correlations
beyond a certain distance between the particles, so a very accurate description of
the critical properties should not be expected. This notwithstanding, as remarked
by Runnels,\cite{runnels:1972} due to the degeneracy of the close-packed
configuration, this system is difficult to study with finite-size or series
expansions method, and a closed-form approximation could be superior at describing
the correct phase behavior.

\subsection{Multicomponent System of Parallel Hard Cubes}

Let us now consider a multicomponent hard cube lattice gas
in a simple cubic lattice. If we denote $\sigma_1,\ldots,
\sigma_p$ the edge-lengths of the different species, then the
interaction potential between species $\alpha$ and $\gamma$, will
be given by
\begin{equation}
\phi_{\alpha \gamma}(\bs,\bs')=\left\{
\begin{array}{ll}
\infty&\text{if $\displaystyle \max_{i=1,2,3}|s_i-s'_i|\leq
\frac{1}{2}(\sigma_\alpha+\sigma_\gamma)$,}\\
0&\text{otherwise.}
\end{array}\right.
\end{equation}
The lattice fundamental measure approximation for the free energy
functional of this system has already been reported in
Ref.~\onlinecite{lafuente:2002b} [Eqs.~(3.2) and (3.3) of
that reference], together with the phase diagram for the
particular case of a binary mixture with
$\sigma_{{\rm L}}=6$ and $\sigma_{{\rm S}}=2$, but no
details about the calculation were given. In this subsection we will
study the phase behavior of the general uniform mixture and obtain
the complete bulk phase diagram, including both uniform and ordered
phases, for the particular case just mentioned.

In the uniform regime, the one-particle distribution function no longer
depends on the spatial variables:
$\rho_\alpha(\bs)=\rho_{\alpha}$, $\alpha=1,\ldots,p$. In this case, 
the free energy density has the simple form\cite{lafuente:2002b}
\begin{multline}\label{fedhom3d}
\Phi(\rho_1,\ldots,\rho_p)=\sum_{\alpha=1}^{p}\rho_\alpha (\ln \rho_\alpha
-1)+\Phi_0(n_3)\\
-\Phi_0(n_2)+\Phi_0(n_1)-\Phi_0(n_0)
\end{multline}
[with the densities $n_k$ defined below Eq.~(\ref{dcf})].

As it is well known, the stability of the mixture is
determined by the matrix
\begin{equation}
\mathsf{M}_{\alpha\gamma}=\frac{\partial^2 \Phi}{\partial \rho_\alpha \partial
\rho_\gamma}.
\end{equation}
In order for the system to be stable in a mixed state this matrix must
be positive definite. As this requirement is
fulfilled in the low density limit,
the spinodal curve can be determined through the equation
$\det \mathsf{M}=0$. When the excess free energy density of the
system depends on the densities only via the finite set of moments
$\xi_l=\sum_{\alpha}\sigma_\alpha^l \rho_\alpha$ ($l=0,\ldots,m$),
the spinodal can be expressed
in the equivalent, but more suitable form\cite{cuesta:1999}
$\det \mathsf{Q}=0$, where
\begin{equation}
\mathsf{Q}_{ij}=\delta_{ij}+\sum_{k=0}^m\xi_{i+k}\Phi_{kj},
\quad
\Phi_{ij}=\frac{\partial^2 \Phi_{{\rm ex}}}{\partial \xi_i \partial \xi_j}.
\end{equation}
This is just our case, because the excess free energy density 
[the $\Phi_0$ contributions in (\ref{fedhom3d})]
depends on the densities through the set $\{n_k\}$, and this variables can
easily be expressed in terms of the set of moments $\{\xi_0,\ldots,\xi_3\}$.
Thus the equation for the spinodal of our system reads
\begin{multline}
(1+2\xi_3)^2-(\xi_1+3\xi_2)(1+2\xi_3)\\-3(\xi_2-\xi_1)(1+\xi_4)
+\xi_2(5\xi_2-\xi_1)=0.
\end{multline}

For a binary mixture with the small component having $\sigma_{{\rm S}}=2$,
it can be shown that the smallest size ratio, $r=\sigma_{{\rm L}}/\sigma_{{\rm S}}$,
necessary to have a spinodal instability is $r=13$.
This value is in strong disagreement with previous simulation
results,\cite{dijkstra:1994} which reported a demixing phase transition
for $r=3$. An explanation of this mismatch has already been provided
in Ref.~\onlinecite{lafuente:2002a}, and will become clear later on.

\begin{figure}
\includegraphics[width=70mm,clip=]{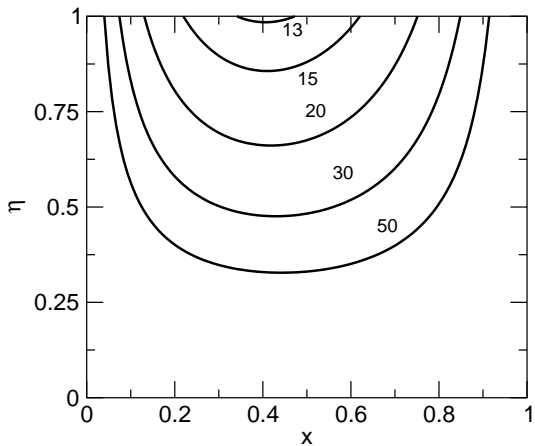}
\caption{\label{fig6}Spinodal curves for a uniform binary mixture with the smallest component
of size $\sigma_{{\rm S}}=2$ and different values of the size ratio,
$r=\sigma_{{\rm L}}/\sigma_{{\rm S}}$.}
\end{figure}

Some spinodals for different size ratios are shown in Fig.~\ref{fig6}.
It should be noticed that the continuum counterpart\cite{martinez-raton:1999}
is recovered in the limit $\sigma_{{\rm S}}\to \infty$, while keeping $r$ constant.
For that system, it was shown that the minimum value of $r$ to find demixing is
$r=5+\sqrt{24}\approx9.98$. Therefore, we can conclude that the lattice enhances the stability
of the mixture. What this analogy with the continuum system
suggests is that we should expect
fluid-fluid demixing to be preempted by the freezing of one of
the coexisting phases also in the lattice model.
It must be remarked that, unlike in the continuum case, in the
lattice system the stability condition involves not only the size ratio,
but also the edge-length of one of the components, thus making the analysis of
the stability more complex.

The discrete nature of this system provides a very suitable framework to study
ordering transitions. In the continuum, only a partial analysis have been done,
\cite{martinez-raton:1999} because the minimization of the functional becomes
a numerically very demanding task. Usually this minimization is performed by restricting
the density
profiles into a parametric class, and then minimizing with respect to
one or a few parameters.
The problem for the mixture is that it is very difficult to guess the
appropriate class. In contrast, the situation
in the lattice is easier to handle because
it is feasible to perform a free minimization with the only constraints 
imposed by the symmetry of the ordered phase and 
its periodicity. As it was shown in the analysis
of the two-dimensional system, the periodicity of the ordered phase
can be estimated from
the divergence of the structure factor,
in the case of a mixture the latter being
a matrix. The analog to the condition (\ref{sf2d}) for the
mixture is
\begin{equation}\label{sf3d}
\det(\mathsf{P}^{-1}-\hat{\mathsf{C}}(\bq))=0,
\end{equation}
where $\mathsf{P}=\big(\delta_{\alpha\gamma}\rho_\alpha\big)$, a
diagonal matrix, and
$\hat{\mathsf{C}}(\bq)=\big(\hat{c}_{\alpha\gamma}(\bq)\big)$
is the matrix of Fourier transforms of the
direct correlation functions between all species.

In the remaining of this subsection
we will restrict ourselves to the particular
case of the binary mixture with $\sigma_{{\rm S}}=2$ and $\sigma_{{\rm L}}=6$,
the only case for which simulations are available.
\cite{dijkstra:1994} The main result of these simulations is that the
mixture undergoes an entropy-driven
fluid-fluid demixing, thus being the only known
example of an athermal additive model showing this feature.

The strategy we have adopted in order to obtain the complete phase diagram for the mixture
has been the following: (i) First, we have calculated the phase diagram for the pure component
systems, both for the small and large particles; (ii) then, we have obtained the
curves marking spatial instabilities
for the whole mixture through
the divergence of the structure factor matrix, and
have calculated
the periodicity of the ordered phases arising at the bifurcation points, and (iii) finally, we have
completed the phase diagram by calculating all the possible phase transitions compatible with the
results obtained in the two previous steps, choosing those thermodynamically
more stable.

For the pure component systems, we will proceed as in the
two-dimensional case.
For $\sigma=2$, the fundamental measure 
excess free energy functional (\ref{functional}) is
given in diagrammatic notation in (\ref{f3d}).
For a general one-component system with edge-length
$\sigma$, the functional form
is obtained particularizing (\ref{functional}), but the
structure is the same as that of (\ref{f3d}).

As it was mentioned in the previous subsection, in a symmetry broken continuous phase
transition condition (\ref{sf2d}) is satisfied.
For $\sigma=2$ it yields
\begin{equation}
\eta_{{\rm crit}}=\sigma^3\rho_{{\rm crit}}=0.568,
\end{equation}
with $q=\pi$ indicating a periodicity $d=2\pi/q=2$,
while for $\sigma=6$ we find the value
\begin{equation}
\eta_{{\rm crit}}=\sigma^3\rho_{{\rm crit}}=0.402,
\end{equation}
and period $d=7$. Notice that at close packing
this system has period $d=6$, so we must consider
both. With respect to the symmetry of the phases, we have
considered the smectic, columnar and solid ones (ordering along one, two and three
coordinate axes, respectively).
\begin{figure}
\includegraphics[width=70mm,clip=]{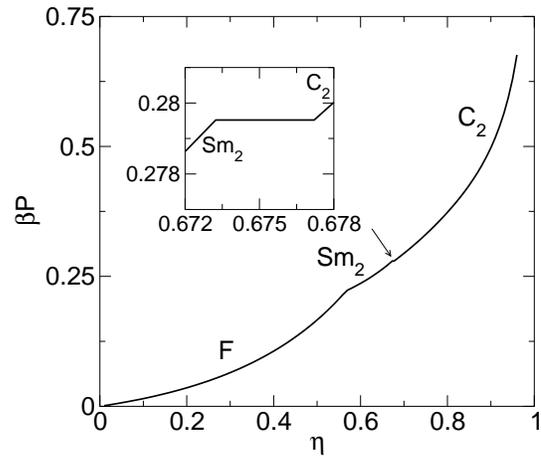}
\caption{\label{fig7}Equation of state (pressure vs.~packing fraction) for a $\sigma=2$
parallel hard cube lattice gas on a simple cubic lattice. The different symmetries are
denoted by F, Sm and C, meaning fluid, smectic and columnar, respectively. The periodicity
has been indicated by a subindex. The inset shows a very narrow first order transition
from a smectic to a columnar phase.}
\end{figure}
\begin{figure}
\includegraphics[width=70mm,clip=]{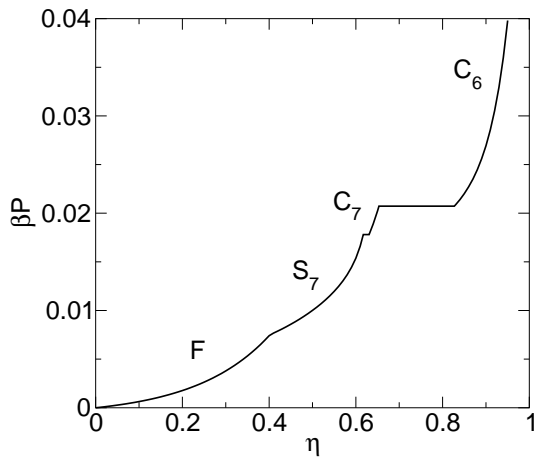}
\caption{\label{fig8}Equation of state (pressure vs.~packing fraction) for a $\sigma=6$
parallel hard cube lattice gas on a simple cubic lattice. The different symmetries are
denoted by F, Sm, C and S, meaning fluid, smectic, columnar and solid, respectively. The periodicity
has been indicated by a subindex.}
\end{figure}

Now it is possible to perform a free minimization of the functional with the above
restrictions. In this case we have to proceed numerically, because the complexity
of the problem does not permit an analytical treatment. This notwithstanding,
the structural form of the functional 
simplifies the numerical work: the
weighted densities are just convolutions which
can be computed by using fast Fourier transform.
To give an idea about the degree of complexity of the problem
we will say that the simplest phase to minimize is the
period-2 smectic, which involves minimization on
two variables, and the most complex one is the
period-7 solid, which involves minimization on
twenty variables.

\begin{figure}
\includegraphics[width=70mm,clip=]{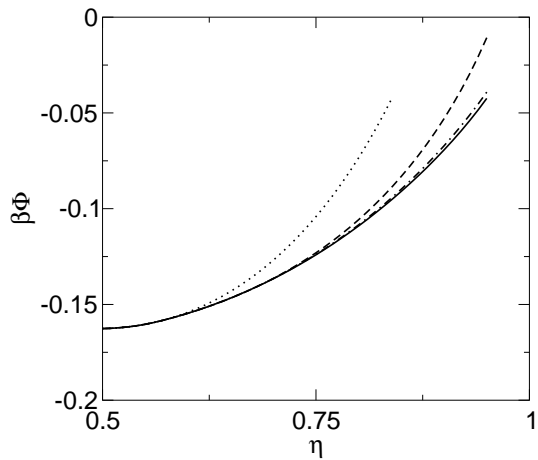}
\caption{\label{fig9} Free energy density of the fluid (dotted line), smectic (dashed line),
columnar (solid line) and solid (dotted-dashed line) phases,
for the system of hard cubes with $\sigma=2$ in a simple cubic lattice.}
\end{figure}

\begin{figure}
\includegraphics[width=70mm,clip=]{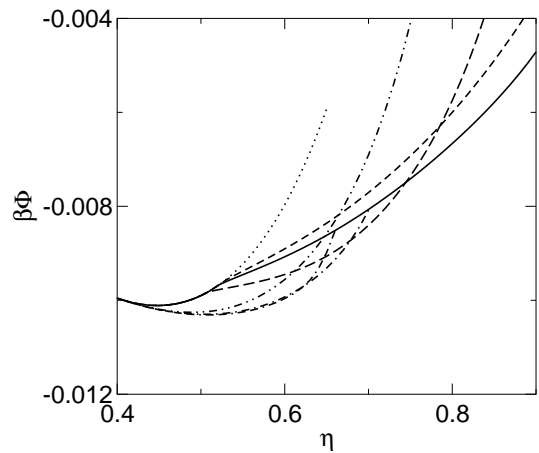}
\caption{\label{fig10} Free energy density of the fluid (dotted line), smectic (double dotted-dashed
line), columnar (dotted-dashed line) and solid (double dashed-dotted line) phases with periodicity $d=7$
and smectic (long-dashed line), columnar (solid line) and solid (dashed line) phases with periodicity $d=6$,
for the same system of the previous figure but with $\sigma=6$.}
\end{figure}

The phase diagrams of both systems, $\sigma=2$ and $\sigma=6$, are shown in Figs.~\ref{fig7}
and \ref{fig8}, respectively. Also, the free energy density near the critical point
appears in Figs.~\ref{fig9} and \ref{fig10}, respectively.
{}From these figures, we can see that there exists a 
strong competition between the different
phases. This reflects in the very narrow first order transitions 
observed in the phase diagrams,
such as the Sm$_2$--C$_2$ coexistence ($\eta_{{\rm Sm}_2}=0.673$ and
$\eta_{{\rm C}_2}=0.677$) in the $\sigma=2$ system, and the S$_7$--C$_7$
($\eta_{{\rm S}_7}=0.617$ and $\eta_{{\rm C}_7}=0.631$) for $\sigma=6$. Since
our treatment is approximate, these phase transitions could 
actually be spurious: given the small differences between the
free energy densities of the phases involved,
other scenarios might be possible. In contrast,
there also exists very well defined transitions
which offer higher confidence, such as the
C$_7$--C$_6$ first order transition ($\eta_{{\rm C}_7}=0.656$ and $\eta_{{\rm C}_6}=0.827$)
in the $\sigma=6$ system.

Let us now consider the binary mixture 
with $\sigma_{{\rm L}}=6$ and $\sigma_{{\rm S}}=2$.
%
\begin{figure}
\includegraphics[width=70mm,clip=]{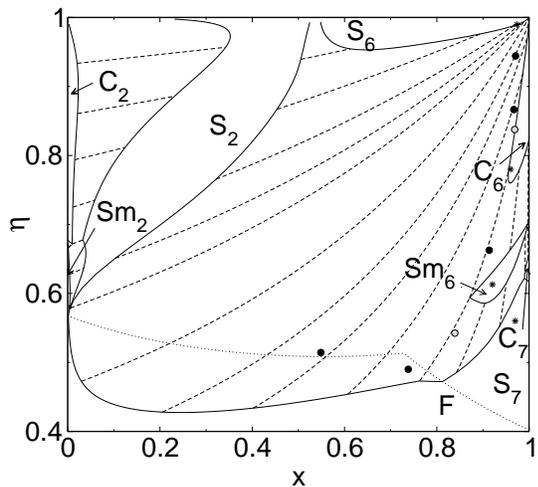}
\caption{\label{fig11} Phase diagram of the binary mixture of
parallel hard cubes (size ratio $6:2$), $\eta=\eta_{{\rm L}}+\eta_{{\rm S}}$
being the total packing fraction of the large (L) and small (S) cubes, and
$x=\eta_{{\rm L}}/\eta$. The phases are labeled F (fluid), Sm$_{\alpha}$ (smectic),
C$_{\alpha}$ (columnar), and S$_{\alpha}$ (solid), where $\alpha=2,6,7$ stands
for the periodicity of the ordered phases. The dashed lines join coexisting states.
The dotted line corresponds to the spinodal of the uniform fluid. For $0.81\lesssim x$
it marks a stable continuous F-S$_7$ phase transition. The circles are coexisting
states taken from the simulation results in Ref.~\onlinecite{dijkstra:1994}. Finally, the stars
correspond to the states whose density profiles are represented in Figs.~\ref{fig13}--\ref{fig16}.} 
\end{figure}
\begin{figure*}
\includegraphics[width=110mm,clip=]{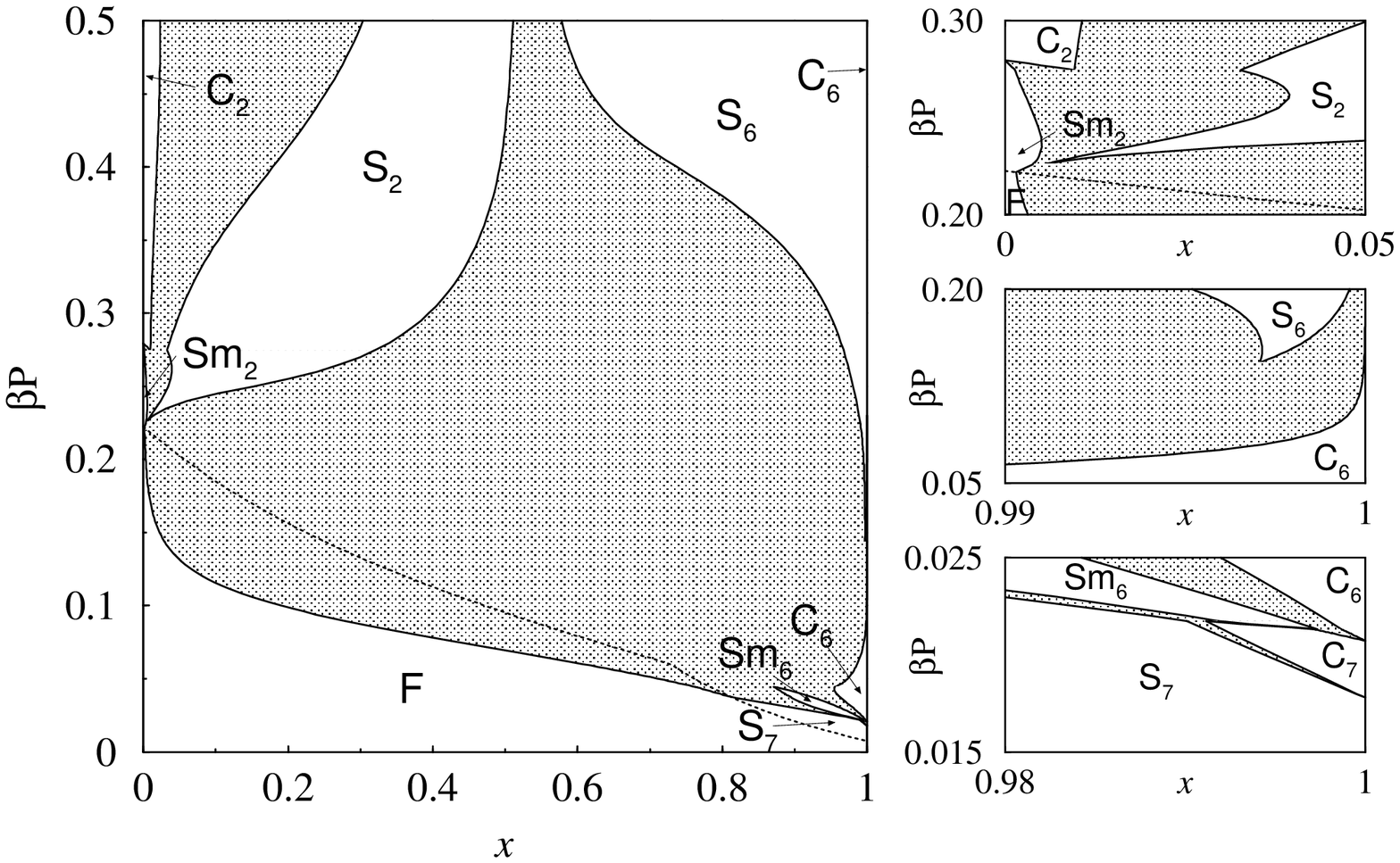}
\caption{\label{fig12} Phase diagram (reduced pressure, $\beta P$, versus composition, $x$)
of the same system refers in Fig.~\ref{fig11} (the labels used are the same as in that figure).
Now coexisting tie lines are horizontal. The dotted line is the same as in the previous figure.
Insets show some details of the phase diagram.}
\end{figure*}
{}From the discussion about the uniform multicomponent
system we can conclude that, for a size ratio $r=3$ and 
a small particle edge-length $\sigma_{{\rm S}}=2$,
there is no fluid-fluid demixing (not even metastable). 
Then, we have to look for spatial instabilities. To
this purpose we must study condition (\ref{sf3d}).
The direct correlation function
is now a $2\times2$ matrix whose elements are given by
(\ref{dcf}).
If we characterize the thermodynamics of our
system by the total packing fraction
$\eta=\eta_{{\rm L}}+\eta_{{\rm S}}$, where
$\eta_{{\rm L(S)}}=\sigma_{{\rm L(S)}}^3\rho_{{\rm L(S)}}$
is the packing fraction of the large (small) cubes, and
by the composition $x=\eta_{{\rm L}}/\eta$,
then, for every value of $x$, we have to look for
the smallest value of $\eta$ which makes the
condition (\ref{sf3d}) solvable. The solution is plotted
in Fig.~\ref{fig11} (Fig.~\ref{fig12} shows the same
phase diagram with $\eta$ replaced by the pressure).
It should be remarked that for $0\leq x\leq 0.728$
the period of the ordered phases
at the fluid spinodal is $d=2$, while for 
$1\geq x\geq 0.728$ we have found $d=7$.

With this guidance we can start looking for the
true phase diagram. This is a very
demanding numerical task, but feasible in a reasonable
amount of time. For each coexistence
curve involving, say, phases P$_1$ and P$_2$, we have to solve the
equilibrium equations
\begin{equation}
\begin{split}
\beta p(\eta_{{\rm P}_1},x_{{\rm P}_1}) &=
\beta p(\eta_{{\rm P}_2},x_{{\rm P}_2}),\\
z_{{\rm L}}(\eta_{{\rm P}_1},x_{{\rm P}_1})&=
z_{{\rm L}}(\eta_{{\rm P}_2},x_{{\rm P}_2}),\\
z_{{\rm S}}(\eta_{{\rm P}_1},x_{{\rm P}_1})&=
z_{{\rm S}}(\eta_{{\rm P}_2},x_{{\rm P}_2}).
\end{split}
\end{equation}
Every iteration of the procedure requires the minimization of the functional, at constant $\eta$ and $x$,
for both phases. In the simplest case this corresponds to a minimization problem with four variables, but
in the most complicated case we have to deal with a forty variable minimization.
Another problem we find is that the subtle
differences between the free energy of different phases,
already encountered in the monocomponent systems, make it very
hard to discern which one is the most stable phase.
So, in many cases it is not clear which coexistence is
thermodynamically more stable. In these doubtful cases, we
have resorted to the Gibbs free energy per
particle, $g(p,X)=X\mu_{{\rm L}}+(1-X)\mu_{{\rm S}}$, where $\mu_{{\rm L,S}}$ is the chemical potential
of each species and $X\equiv\rho_{\rm L}/\rho=
x/(r+(1-r)x)$ is the molar fraction. When $g(p,X)$ is plotted at constant
pressure as a function of $X$, the coexisting phases
in the mixture can be found through a
double tangent construction.

\begin{figure}
\mbox{
\subfigure[Large particles]{
\includegraphics[width=70mm,clip=]{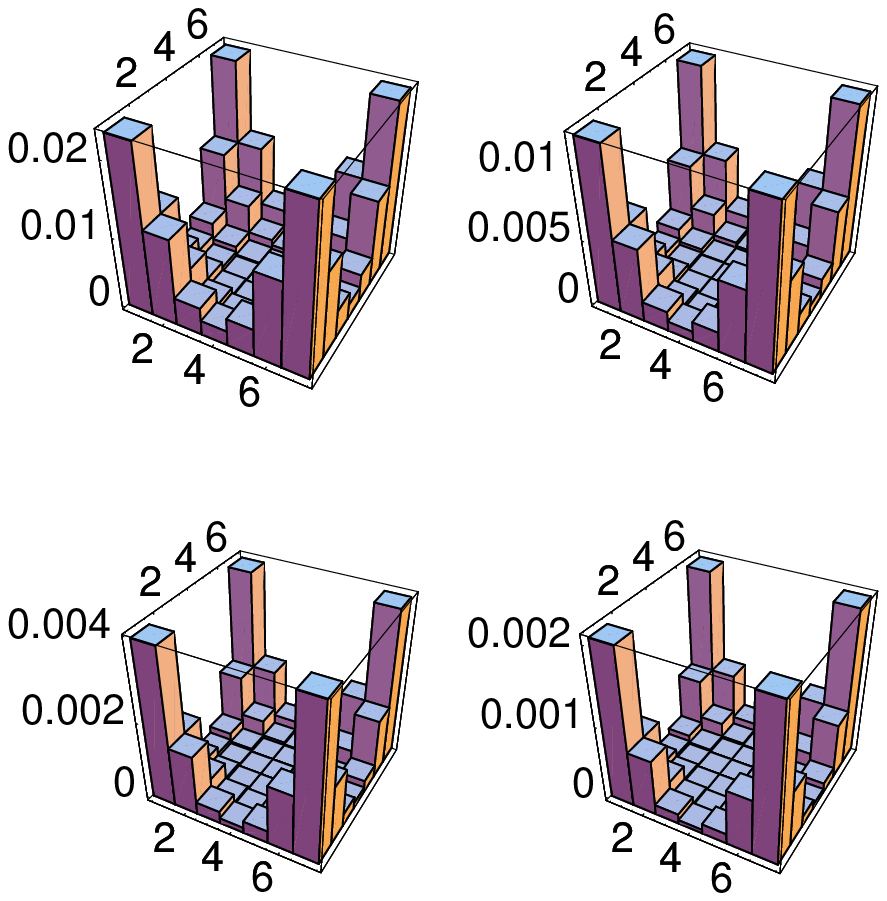}}}
\mbox{
\subfigure[Small particles]{
\includegraphics[width=70mm,clip=]{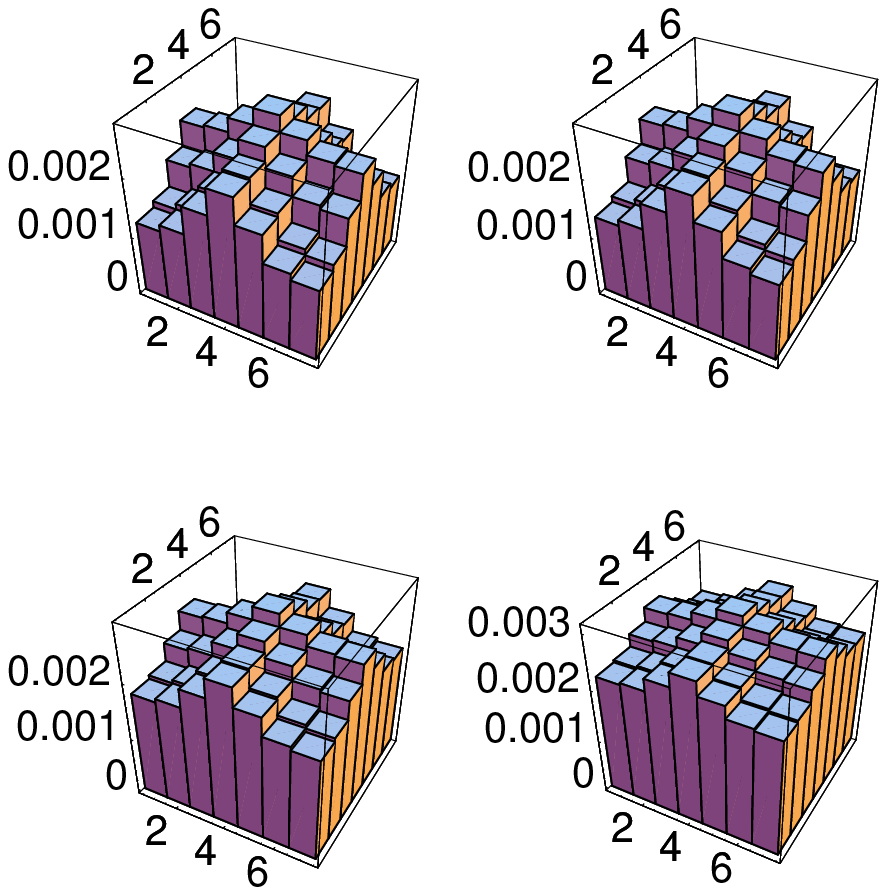}}}
\caption{\label{fig13} Density profiles for a solid phase
with periodicity $d=7$, corresponding to $\eta=0.56$,
$x=0.97$ and $\beta p=0.02$. Different planar sections
at $s_3=1,2,3,4$ (from top to bottom and from
left to right) are plotted.}
\end{figure}

\begin{figure}
\mbox{
\subfigure[Large particles]{
\includegraphics[width=70mm,clip=]{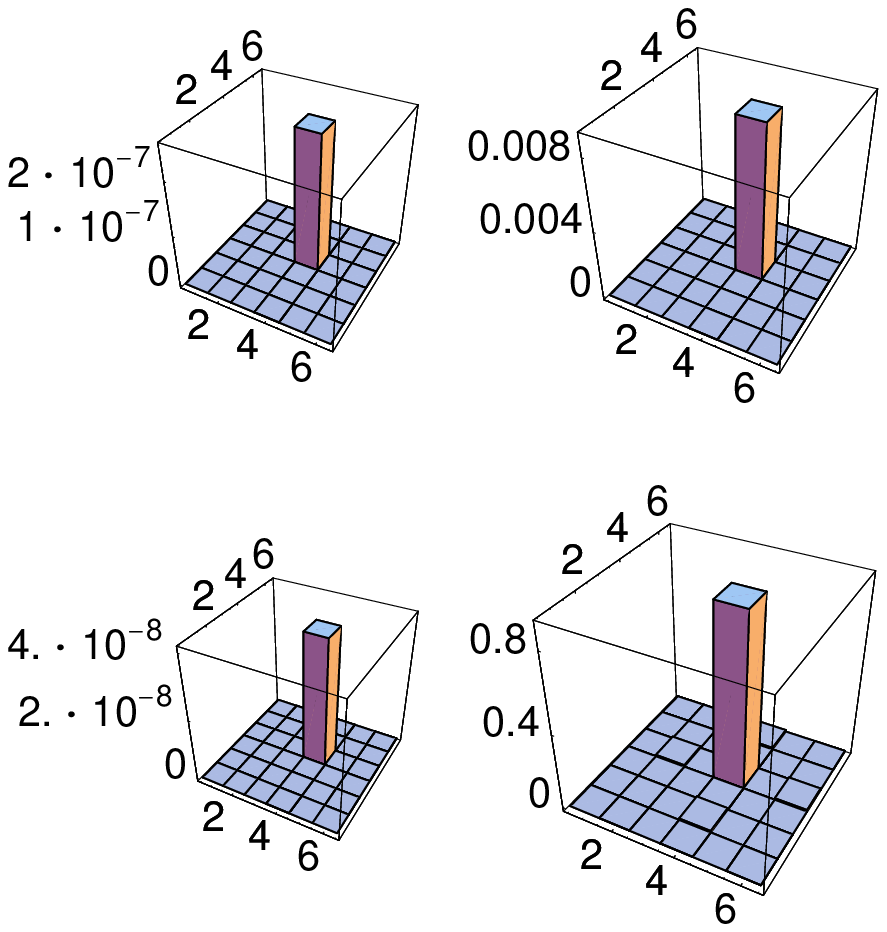}}}
\mbox{
\subfigure[Small particles]{
\includegraphics[width=70mm,clip=]{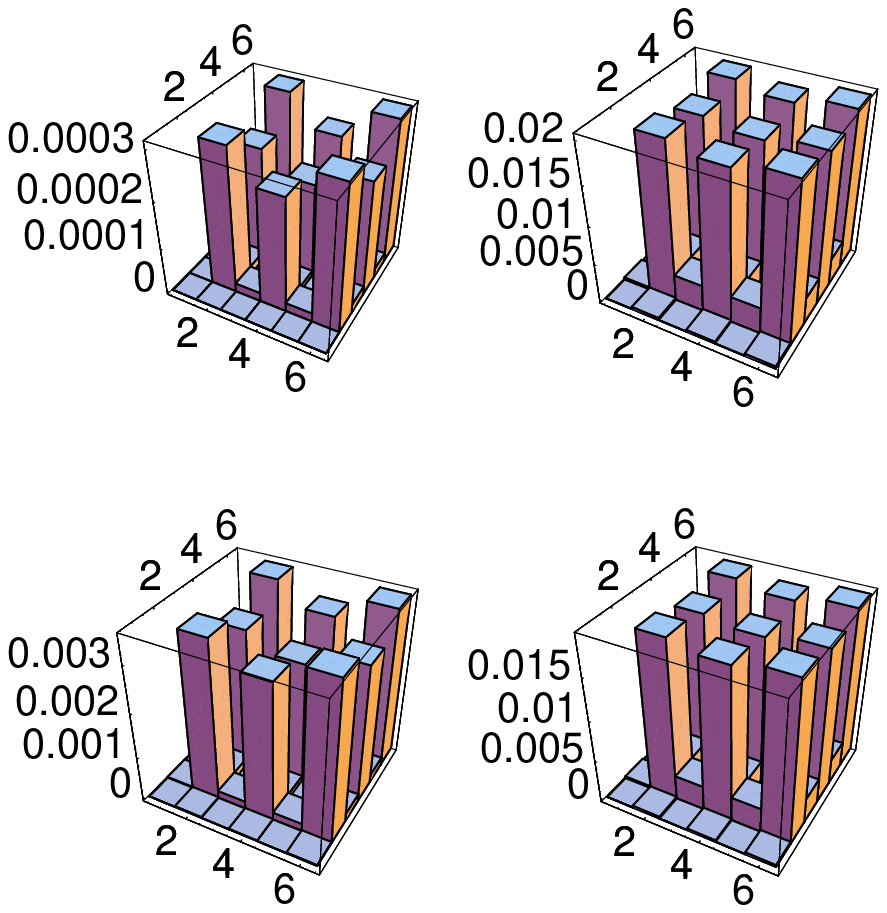}}}
\caption{\label{fig14} Density profiles for a solid phase
with periodicity $d=6$, corresponding to $\eta=0.989$,
$x=0.973$ and $\beta p=0.265$. Different planar sections
at $s_3=1,2,3,4$ (from top to bottom and from
left to right) are plotted.}
\end{figure}

The complete phase diagram appears in Figs.~\ref{fig11}
and \ref{fig12} in two different representations.
In Fig.~\ref{fig12}, it can be observed (see the insets)
that there exist very small parts on
the phase diagram with a plethora of very narrow coexistence
regions. As in the monocomponent case, many of them could just be
spurious. One of the most remarkable features is that there exists
a wide phase separation between a small-particle-rich fluid phase
and a large-particle-rich columnar
phase (which becomes a solid phase for higher pressures).
As explained in the Introduction, this is the usual scenario
for this kind of mixtures. The revision of the simulations
(also shown in Fig.~\ref{fig11}) 
resulting from this phase behavior
has already been discussed in detail
in Ref.~\onlinecite{lafuente:2002a}. The main consequence of this
comparison is that some of the state points obtained in the
simulations must have been misinterpreted as a fluid, while
they should exhibit columnar ordering. Another interesting
result is that, at approximately $\beta p\approx0.24$, there
appear a solid-solid phase separation between a small-particle-rich
and a large-particle-rich phases. Finally, it is worth
mentioning the existence of an extremely narrow
chimney of S$_6$--C$_6$ coexistence. This suggests
that the columnar phase is very sensitive to small
perturbations.

\begin{figure}
\mbox{
\subfigure[Large particles]{
\includegraphics[width=35mm,clip=]{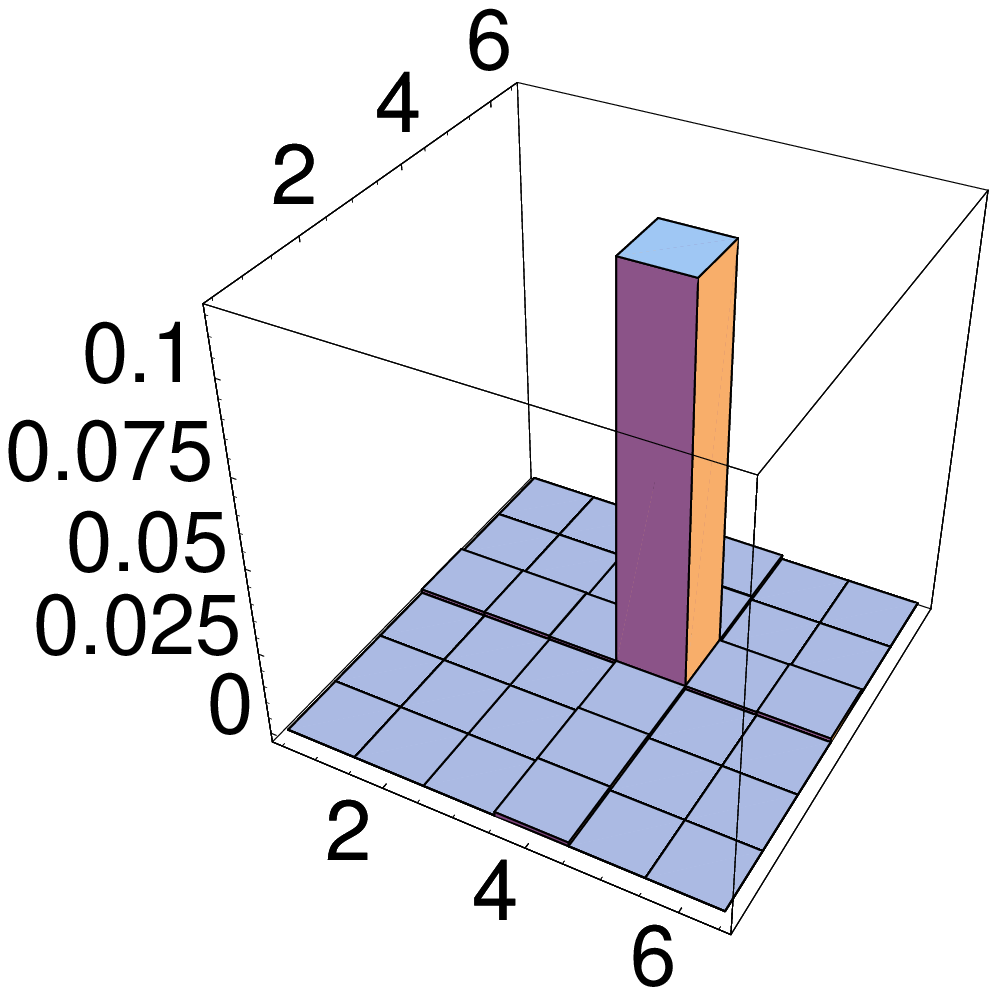}}\hspace{10mm}
\subfigure[Small particles]{
\includegraphics[width=35mm,clip=]{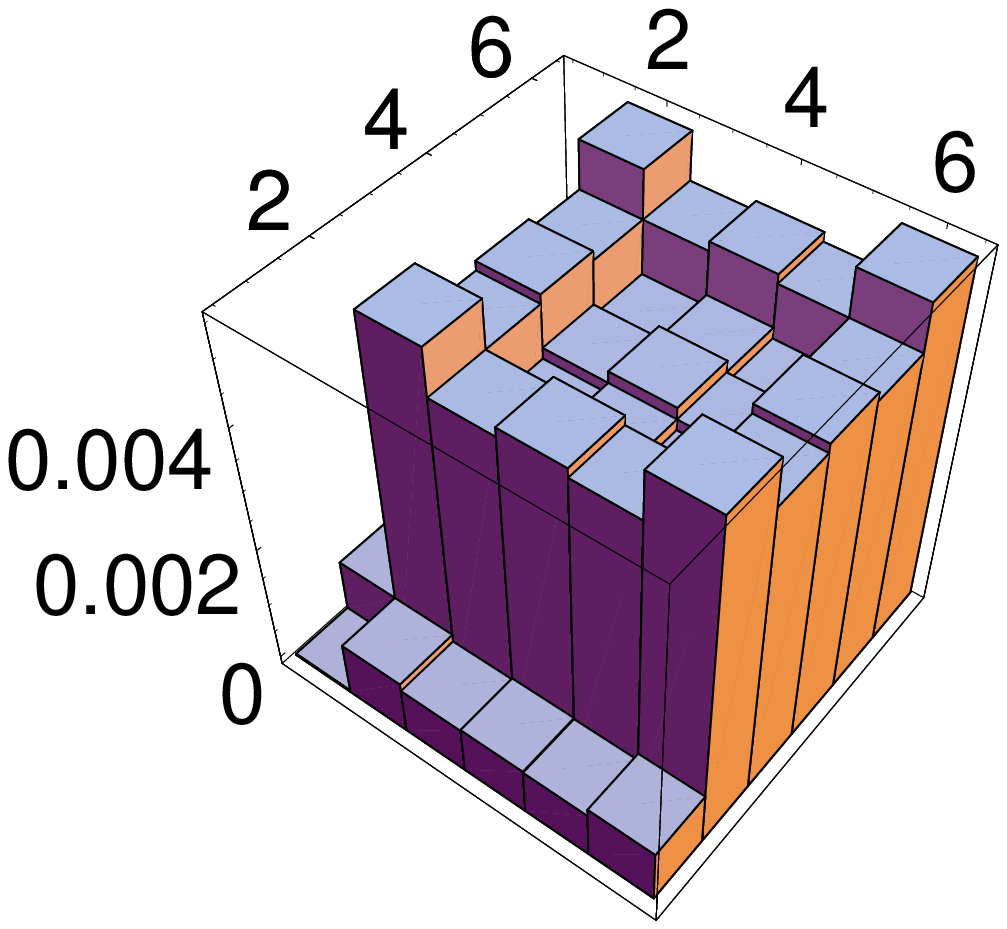}}}
\caption{\label{fig15} Density profiles for a columnar phase
with periodicity $d=6$, corresponding to $\eta=0.78$,
$x=0.96$ and $\beta p=0.04$.}
\end{figure}

\begin{figure}
\mbox{
\subfigure[Large particles]{
\includegraphics[width=35mm,clip=]{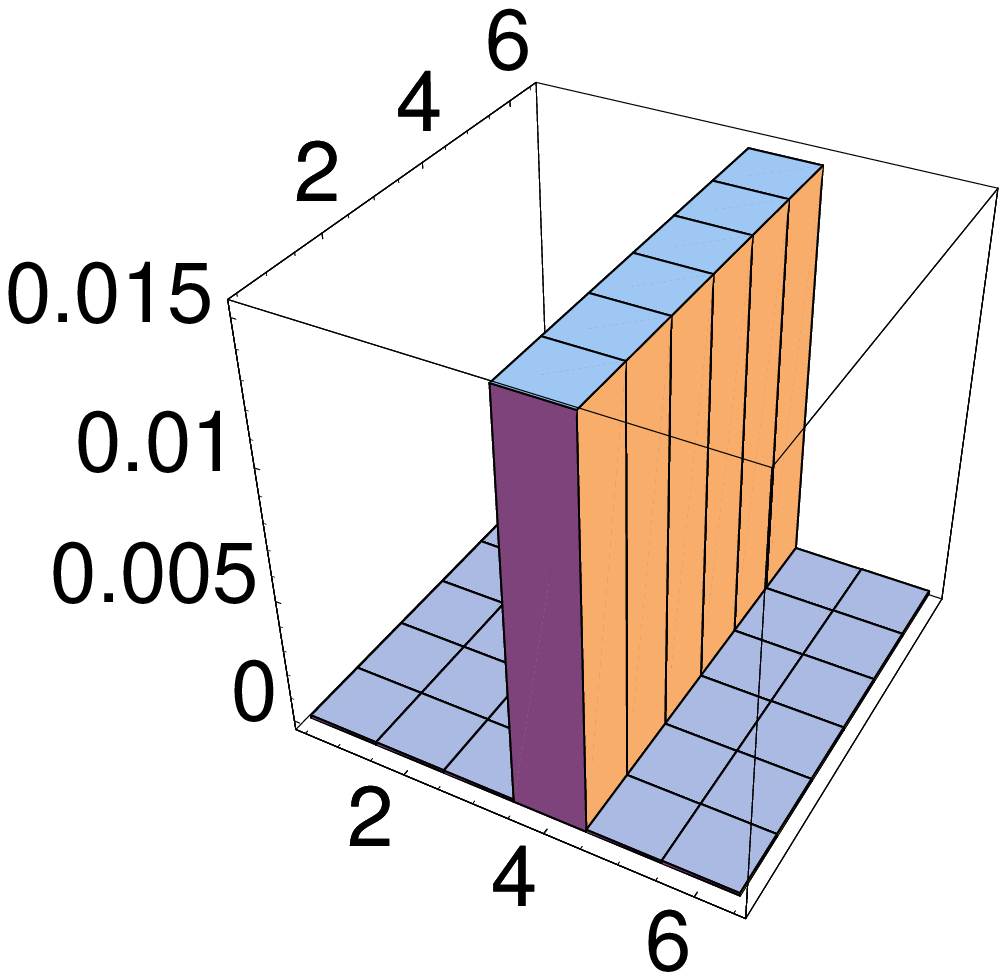}}\hspace{10mm}
\subfigure[Small particles]{
\includegraphics[width=35mm,clip=]{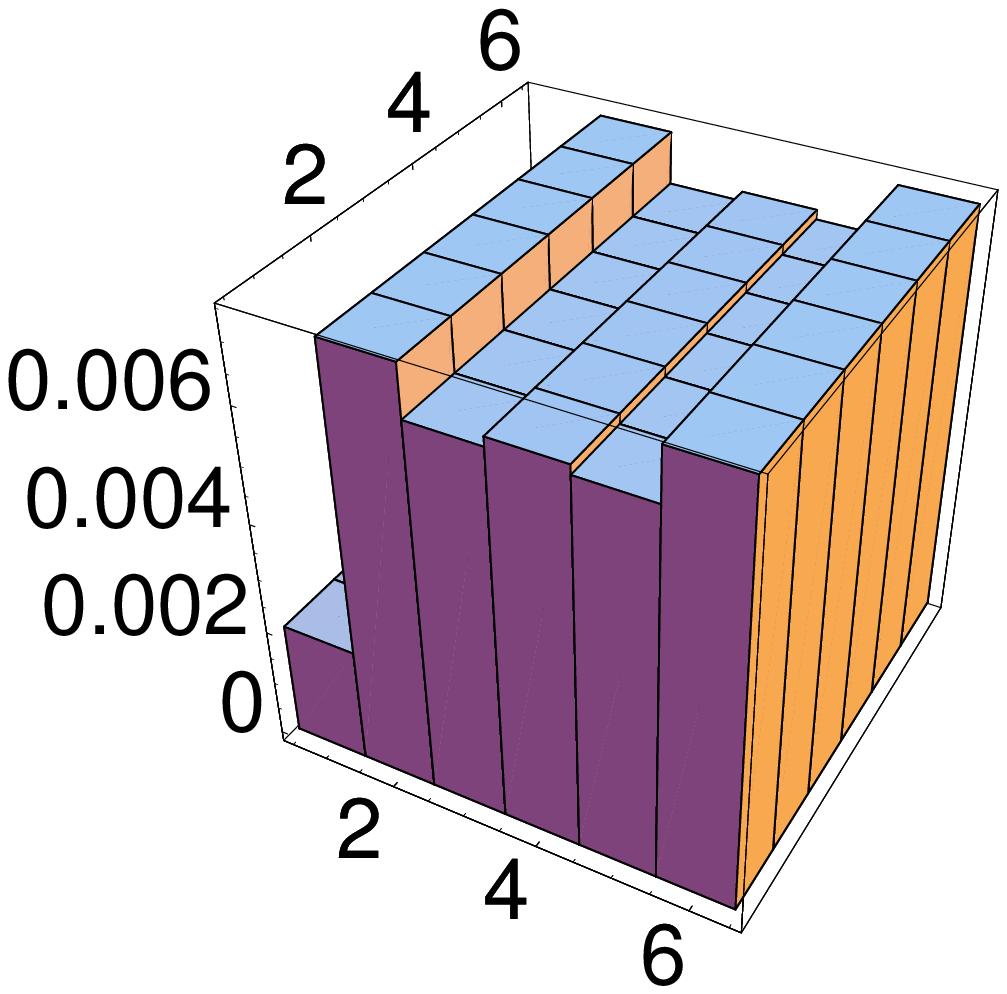}}}
\caption{\label{fig16} Density profiles for a smectic phase
with periodicity $d=6$, corresponding to $\eta=0.92$,
$x=0.613$ and $\beta p=0.035$.}
\end{figure}

As concerns the density profile in the ordered phases,
we have chosen a few representative states of the
system 
(marked with stars in Fig.~\ref{fig11})
in order to illustrate the way large and
small particles distribute in each phase.
In Fig.~\ref{fig13} we show the density profiles
for a period-7 solid phase for different
sections perpendicular to the direction $s_3$. This
size is one lattice spacing bigger than the size
of the large particles. We can conclude from
the figure that the large particles are arranged in
such a way that the small ones can be accommodated between any
two of them. So, we could say that the large particles
in the unit cell are surrounded by the small ones. Note
that the density of the latter does not change very much
within the unit cell. In Fig.~\ref{fig14} the density
profile for a period-6 solid
is shown. It is very interesting
to see that its structure is completely different to the
previous one. Now, as the unit cell is of the same size
of the large particles, the small ones can only be placed
at vacant unit cells. Thus,
the unit cell is completely filled with the small
particles, which, as can be appreciated in the figure,
form a crystal. Noticeably,
the value of the small particle density is
slightly higher at the contact with the large particles,
which can be interpreted as an adsorption phenomenon.
Figures \ref{fig15} and \ref{fig16} exhibit the profile
of the period-6 column and smectic phases, respectively.
{}From the discussion of the period-6 solid profiles
we have just made, the interpretation of these two
new density profiles should be straightforward.

\section{Conclusions}
\label{conclusions}

In this work we have applied the extension
of FMT to lattice models\cite{lafuente:2002b}
to study two systems already treated in the literature: the 
two-dimensional lattice gas with first
and second neighbor exclusion on a square
lattice, and the binary mixture of parallel hard
cubes with edge-lengths
$\sigma_{{\rm L}}=6$ and $\sigma_{{\rm S}}=2$
on a simple cubic lattice.
For both systems we have shown a 
systematic way of using the theory in order to perform a complete calculation of
the bulk phase diagrams. In particular, for the first system a very detailed analysis
have been carried out. All the relevant thermodynamic functions can be analytically
obtained within this approach, for both uniform and columnar phases.
The results compare well with others previously obtained
by different authors using widely accepted theoretical approaches.
It is remarkable that,
in spite of the simplicity of the treatment,
the results obtained are rather accurate, specially in
the low and high density limits.

On the other hand, a very complete study has been carried
out for the three-dimensional system.
The rich phase diagram obtained is very
striking, considering the simplicity
of both the system and the theory. There appear many different
entropy-driven phase transitions: fluid-ordered phase
demixing, one-, two- and three-dimensional ordering and
solid-solid phase separation. 
The results obtained have also allowed to reinterpret
the simulations results\cite{dijkstra:1994}
in a way consistent with the general picture emerged
during the last decade about demixing of additive hard-core
binary mixtures: fluid-fluid demixing is always
preempted by the ordering of one of the phases.

We have performed a free minimization
of the functional and have thus obtained
the structure of all the ordered phases. The results we get show
that the density profile of the small particles is far from being
uniform. In general, when dealing with three-dimensional
models, a free minimization is not feasible, so the density
profile must be properly parametrized. Since there is very little
intuition about the density profiles of mixtures, the small
components are always assumed to be uniformly distributed
over the volume. According to our findings, this is 
definitely wrong. Perhaps our results can help to gain
insight into what a proper parametrization of the density
profiles looks like.

We would also like to emphasize that the LFMT is
a mean-field-like theory, but not a trivial one.
This can be understood if we realize that the direct correlation
functions have finite range, which means that at some point
the correlations between particles are neglected.
This fact is reflected in the lower accuracy
of the description around critical points. However, away
from these regions the results are far more accurate
than those obtained from a standard
mean field theory, and what is even more important,
at the expese of no much more work.

The study
of lattice gases within the framework of density functional
theory can be very fruitful, as we hope to have been
able to transmit in this work. On the one hand, there is 
not loss in phenomenology or complexity of behaviors;
on the other hand, the approach is considerably simpler
numerically, something that allows one to tackle problems
which have so far not been solved in continuum models. 
Besides, from a purely formal point of view, lattice density
functionals reveal some structures which may be hidden in
similar developments for continuum models. So, its 
careful analysis may reveal important properties of the
latter in the near future. For these reasons we believe
that it is very important to extend this kind of work
to more general lattice models. We plan to report on that
shortly.

\begin{acknowledgments}
This work is supported by project BFM2000-0004 from the Direcci\'on
General de Investigaci\'on (DGI) of the Spanish Ministerio de Ciencia
y Tecnolog\'{\i}a.
\end{acknowledgments}

\bibliography{llafuent}

\begin{thebibliography}{55}
\expandafter\ifx\csname natexlab\endcsname\relax\def\natexlab#1{#1}\fi
\expandafter\ifx\csname bibnamefont\endcsname\relax
  \def\bibnamefont#1{#1}\fi
\expandafter\ifx\csname bibfnamefont\endcsname\relax
  \def\bibfnamefont#1{#1}\fi
\expandafter\ifx\csname citenamefont\endcsname\relax
  \def\citenamefont#1{#1}\fi
\expandafter\ifx\csname url\endcsname\relax
  \def\url#1{\texttt{#1}}\fi
\expandafter\ifx\csname urlprefix\endcsname\relax\def\urlprefix{URL }\fi
\providecommand{\bibinfo}[2]{#2}
\providecommand{\eprint}[2][]{\url{#2}}

\bibitem[{\citenamefont{Onsager}(1949)}]{onsager:1949}
\bibinfo{author}{\bibfnamefont{L.}~\bibnamefont{Onsager}},
  \bibinfo{journal}{Ann. N.Y. Acad. Sci.} \textbf{\bibinfo{volume}{51}},
  \bibinfo{pages}{627} (\bibinfo{year}{1949}).

\bibitem[{\citenamefont{Kirkwood et~al.}(1950)\citenamefont{Kirkwood, Maun, and
  Alder}}]{kirkwood:1950}
\bibinfo{author}{\bibfnamefont{J.~G.} \bibnamefont{Kirkwood}},
  \bibinfo{author}{\bibfnamefont{E.~K.} \bibnamefont{Maun}}, \bibnamefont{and}
  \bibinfo{author}{\bibfnamefont{B.~J.} \bibnamefont{Alder}},
  \bibinfo{journal}{J. Chem. Phys.} \textbf{\bibinfo{volume}{18}},
  \bibinfo{pages}{1040} (\bibinfo{year}{1950}).

\bibitem[{\citenamefont{Alder and Wainwright}(1957)}]{alder:1957}
\bibinfo{author}{\bibfnamefont{B.~J.} \bibnamefont{Alder}} \bibnamefont{and}
  \bibinfo{author}{\bibfnamefont{T.~E.} \bibnamefont{Wainwright}},
  \bibinfo{journal}{J.\ Chem.\ Phys.} \textbf{\bibinfo{volume}{27}},
  \bibinfo{pages}{1208} (\bibinfo{year}{1957}).

\bibitem[{\citenamefont{Wood and Jacobson}(1957)}]{wood:1957}
\bibinfo{author}{\bibfnamefont{W.~W.} \bibnamefont{Wood}} \bibnamefont{and}
  \bibinfo{author}{\bibfnamefont{J.~D.} \bibnamefont{Jacobson}},
  \bibinfo{journal}{J.\ Chem.\ Phys.} \textbf{\bibinfo{volume}{27}},
  \bibinfo{pages}{1207} (\bibinfo{year}{1957}).

\bibitem[{\citenamefont{Frenkel et~al.}(1988)\citenamefont{Frenkel,
  Lekkerkerker, and Stroobants}}]{frenkel:1988}
\bibinfo{author}{\bibfnamefont{D.}~\bibnamefont{Frenkel}},
  \bibinfo{author}{\bibfnamefont{H.~N.~V.} \bibnamefont{Lekkerkerker}},
  \bibnamefont{and}
  \bibinfo{author}{\bibfnamefont{A.}~\bibnamefont{Stroobants}},
  \bibinfo{journal}{Nature} \textbf{\bibinfo{volume}{332}},
  \bibinfo{pages}{822} (\bibinfo{year}{1988}).

\bibitem[{\citenamefont{Veerman and Frenkel}(1992)}]{veerman:1992}
\bibinfo{author}{\bibfnamefont{J.~A.~C.} \bibnamefont{Veerman}}
  \bibnamefont{and} \bibinfo{author}{\bibfnamefont{D.}~\bibnamefont{Frenkel}},
  \bibinfo{journal}{Phys. Rev. A} \textbf{\bibinfo{volume}{45}},
  \bibinfo{pages}{5633} (\bibinfo{year}{1992}).

\bibitem[{\citenamefont{Bolhuis and Frenkel}(1997)}]{bolhuis:1997}
\bibinfo{author}{\bibfnamefont{P.}~\bibnamefont{Bolhuis}} \bibnamefont{and}
  \bibinfo{author}{\bibfnamefont{D.}~\bibnamefont{Frenkel}},
  \bibinfo{journal}{J. Chem. Phys.} \textbf{\bibinfo{volume}{106}},
  \bibinfo{pages}{666} (\bibinfo{year}{1997}).

\bibitem[{\citenamefont{Widom and Rowlinson}(1970)}]{widom:1970}
\bibinfo{author}{\bibfnamefont{B.}~\bibnamefont{Widom}} \bibnamefont{and}
  \bibinfo{author}{\bibfnamefont{J.~S.} \bibnamefont{Rowlinson}},
  \bibinfo{journal}{J.\ Chem.\ Phys.} \textbf{\bibinfo{volume}{52}},
  \bibinfo{pages}{1670} (\bibinfo{year}{1970}).

\bibitem[{\citenamefont{Widom}(1967)}]{widom:1967}
\bibinfo{author}{\bibfnamefont{B.}~\bibnamefont{Widom}}, \bibinfo{journal}{J.\
  Chem.\ Phys.} \textbf{\bibinfo{volume}{46}}, \bibinfo{pages}{3324}
  (\bibinfo{year}{1967}).

\bibitem[{\citenamefont{Frenkel and Louis}(1992)}]{frenkel:1992}
\bibinfo{author}{\bibfnamefont{D.}~\bibnamefont{Frenkel}} \bibnamefont{and}
  \bibinfo{author}{\bibfnamefont{A.}~\bibnamefont{Louis}},
  \bibinfo{journal}{Phys.\ Rev.\ Lett.} \textbf{\bibinfo{volume}{68}},
  \bibinfo{pages}{3363} (\bibinfo{year}{1992}).

\bibitem[{\citenamefont{Asakura and Oosawa}(1954)}]{asakura:1954}
\bibinfo{author}{\bibfnamefont{S.}~\bibnamefont{Asakura}} \bibnamefont{and}
  \bibinfo{author}{\bibfnamefont{F.}~\bibnamefont{Oosawa}},
  \bibinfo{journal}{J.\ Chem.\ Phys.} \textbf{\bibinfo{volume}{22}},
  \bibinfo{pages}{1255} (\bibinfo{year}{1954}).

\bibitem[{\citenamefont{Lebowitz and Rowlinson}(1964)}]{lebowitz:1964}
\bibinfo{author}{\bibfnamefont{J.~L.} \bibnamefont{Lebowitz}} \bibnamefont{and}
  \bibinfo{author}{\bibfnamefont{J.~S.} \bibnamefont{Rowlinson}},
  \bibinfo{journal}{J.\ Chem.\ Phys.} \textbf{\bibinfo{volume}{41}},
  \bibinfo{pages}{133} (\bibinfo{year}{1964}).

\bibitem[{\citenamefont{Biben and Hansen}(1991)}]{biben:1991}
\bibinfo{author}{\bibfnamefont{T.}~\bibnamefont{Biben}} \bibnamefont{and}
  \bibinfo{author}{\bibfnamefont{J.-P.} \bibnamefont{Hansen}},
  \bibinfo{journal}{Phys.\ Rev.\ Lett.} \textbf{\bibinfo{volume}{66}},
  \bibinfo{pages}{2215} (\bibinfo{year}{1991}).

\bibitem[{\citenamefont{Lekkerkerker and Stroobants}(1993)}]{lekkerkerker:1993}
\bibinfo{author}{\bibfnamefont{H.~N.~W.} \bibnamefont{Lekkerkerker}}
  \bibnamefont{and}
  \bibinfo{author}{\bibfnamefont{A.}~\bibnamefont{Stroobants}},
  \bibinfo{journal}{Physica A} \textbf{\bibinfo{volume}{195}},
  \bibinfo{pages}{387} (\bibinfo{year}{1993}).

\bibitem[{\citenamefont{Rosenfeld}(1994)}]{rosenfeld:1994}
\bibinfo{author}{\bibfnamefont{Y.}~\bibnamefont{Rosenfeld}},
  \bibinfo{journal}{Phys.\ Rev.\ Lett.} \textbf{\bibinfo{volume}{72}},
  \bibinfo{pages}{3831} (\bibinfo{year}{1994}).

\bibitem[{\citenamefont{Poon and Warren}(1994)}]{poon:1994}
\bibinfo{author}{\bibfnamefont{W.~C.~K.} \bibnamefont{Poon}} \bibnamefont{and}
  \bibinfo{author}{\bibfnamefont{P.~B.} \bibnamefont{Warren}},
  \bibinfo{journal}{Europhys.\ Lett.} \textbf{\bibinfo{volume}{28}},
  \bibinfo{pages}{513} (\bibinfo{year}{1994}).

\bibitem[{\citenamefont{Caccamo and Pellicane}(1997)}]{caccamo:1997}
\bibinfo{author}{\bibfnamefont{C.}~\bibnamefont{Caccamo}} \bibnamefont{and}
  \bibinfo{author}{\bibfnamefont{G.}~\bibnamefont{Pellicane}},
  \bibinfo{journal}{Physica A} \textbf{\bibinfo{volume}{235}},
  \bibinfo{pages}{149} (\bibinfo{year}{1997}).

\bibitem[{\citenamefont{Coussaert and Baus}(1998)}]{coussaert:1998}
\bibinfo{author}{\bibfnamefont{T.}~\bibnamefont{Coussaert}} \bibnamefont{and}
  \bibinfo{author}{\bibfnamefont{M.}~\bibnamefont{Baus}}, \bibinfo{journal}{J.
  Chem. Phys.} \textbf{\bibinfo{volume}{109}}, \bibinfo{pages}{6012}
  (\bibinfo{year}{1998}).

\bibitem[{\citenamefont{Velasco et~al.}(1999)\citenamefont{Velasco,
  Navascu\'es, and Mederos}}]{velasco:1999}
\bibinfo{author}{\bibfnamefont{E.}~\bibnamefont{Velasco}},
  \bibinfo{author}{\bibfnamefont{G.}~\bibnamefont{Navascu\'es}},
  \bibnamefont{and} \bibinfo{author}{\bibfnamefont{L.}~\bibnamefont{Mederos}},
  \bibinfo{journal}{Phys. Rev. E} \textbf{\bibinfo{volume}{60}},
  \bibinfo{pages}{3158} (\bibinfo{year}{1999}).

\bibitem[{\citenamefont{Buhot and Krauth}(1998)}]{buhot:1998}
\bibinfo{author}{\bibfnamefont{A.}~\bibnamefont{Buhot}} \bibnamefont{and}
  \bibinfo{author}{\bibfnamefont{W.}~\bibnamefont{Krauth}},
  \bibinfo{journal}{Phys.\ Rev.\ Lett.} \textbf{\bibinfo{volume}{80}},
  \bibinfo{pages}{3787} (\bibinfo{year}{1998}).

\bibitem[{\citenamefont{Dijkstra et~al.}(1998)\citenamefont{Dijkstra, van Roij,
  and Evans}}]{dijkstra:1998a}
\bibinfo{author}{\bibfnamefont{M.}~\bibnamefont{Dijkstra}},
  \bibinfo{author}{\bibfnamefont{R.}~\bibnamefont{van Roij}}, \bibnamefont{and}
  \bibinfo{author}{\bibfnamefont{R.}~\bibnamefont{Evans}},
  \bibinfo{journal}{Phys.\ Rev.\ Lett.} \textbf{\bibinfo{volume}{81}},
  \bibinfo{pages}{2268} (\bibinfo{year}{1998}).

\bibitem[{\citenamefont{Dijkstra et~al.}(1999)\citenamefont{Dijkstra, van Roij,
  and Evans}}]{dijkstra:1999}
\bibinfo{author}{\bibfnamefont{M.}~\bibnamefont{Dijkstra}},
  \bibinfo{author}{\bibfnamefont{R.}~\bibnamefont{van Roij}}, \bibnamefont{and}
  \bibinfo{author}{\bibfnamefont{R.}~\bibnamefont{Evans}},
  \bibinfo{journal}{Phys. Rev. Lett.} \textbf{\bibinfo{volume}{82}},
  \bibinfo{pages}{117} (\bibinfo{year}{1999}).

\bibitem[{\citenamefont{Garc\'{\i}a-Almarza and
  Enciso}(1999)}]{garcia-almarza:1999}
\bibinfo{author}{\bibfnamefont{N.}~\bibnamefont{Garc\'{\i}a-Almarza}}
  \bibnamefont{and} \bibinfo{author}{\bibfnamefont{E.}~\bibnamefont{Enciso}},
  \bibinfo{journal}{Phys.\ Rev.\ E} \textbf{\bibinfo{volume}{59}},
  \bibinfo{pages}{4426} (\bibinfo{year}{1999}).

\bibitem[{\citenamefont{Kaplan et~al.}(1994)\citenamefont{Kaplan, Rouke, Yodh,
  and Pine}}]{kaplan:1994}
\bibinfo{author}{\bibfnamefont{P.~D.} \bibnamefont{Kaplan}},
  \bibinfo{author}{\bibfnamefont{J.~L.} \bibnamefont{Rouke}},
  \bibinfo{author}{\bibfnamefont{A.~G.} \bibnamefont{Yodh}}, \bibnamefont{and}
  \bibinfo{author}{\bibfnamefont{D.~J.} \bibnamefont{Pine}},
  \bibinfo{journal}{Phys.\ Rev.\ Lett.} \textbf{\bibinfo{volume}{72}},
  \bibinfo{pages}{582} (\bibinfo{year}{1994}).

\bibitem[{\citenamefont{Dinsmore et~al.}(1995)\citenamefont{Dinsmore, Yodh, and
  Pine}}]{dinsmore:1995}
\bibinfo{author}{\bibfnamefont{A.~D.} \bibnamefont{Dinsmore}},
  \bibinfo{author}{\bibfnamefont{A.~G.} \bibnamefont{Yodh}}, \bibnamefont{and}
  \bibinfo{author}{\bibfnamefont{D.~J.} \bibnamefont{Pine}},
  \bibinfo{journal}{Phys.\ Rev.\ E} \textbf{\bibinfo{volume}{52}},
  \bibinfo{pages}{4045} (\bibinfo{year}{1995}).

\bibitem[{\citenamefont{Steiner et~al.}(1995)\citenamefont{Steiner, Meller, and
  Stavans}}]{steiner:1995}
\bibinfo{author}{\bibfnamefont{U.}~\bibnamefont{Steiner}},
  \bibinfo{author}{\bibfnamefont{A.}~\bibnamefont{Meller}}, \bibnamefont{and}
  \bibinfo{author}{\bibfnamefont{J.}~\bibnamefont{Stavans}},
  \bibinfo{journal}{Phys.\ Rev.\ Lett.} \textbf{\bibinfo{volume}{74}},
  \bibinfo{pages}{4750} (\bibinfo{year}{1995}).

\bibitem[{\citenamefont{Imhof and Dhont}(1995)}]{imhof:1995}
\bibinfo{author}{\bibfnamefont{A.}~\bibnamefont{Imhof}} \bibnamefont{and}
  \bibinfo{author}{\bibfnamefont{J.~K.~G.} \bibnamefont{Dhont}},
  \bibinfo{journal}{Phys.\ Rev.\ Lett.} \textbf{\bibinfo{volume}{75}},
  \bibinfo{pages}{1662} (\bibinfo{year}{1995}).

\bibitem[{\citenamefont{Mart\'{\i}nez-Rat\'on and
  Cuesta}(1999)}]{martinez-raton:1999}
\bibinfo{author}{\bibfnamefont{Y.}~\bibnamefont{Mart\'{\i}nez-Rat\'on}}
  \bibnamefont{and} \bibinfo{author}{\bibfnamefont{J.~A.}
  \bibnamefont{Cuesta}}, \bibinfo{journal}{J.\ Chem.\ Phys.}
  \textbf{\bibinfo{volume}{111}}, \bibinfo{pages}{317} (\bibinfo{year}{1999}).

\bibitem[{\citenamefont{Cuesta}(1996)}]{cuesta:1996}
\bibinfo{author}{\bibfnamefont{J.~A.} \bibnamefont{Cuesta}},
  \bibinfo{journal}{Phys.\ Rev.\ Lett.} \textbf{\bibinfo{volume}{76}},
  \bibinfo{pages}{3742} (\bibinfo{year}{1996}).

\bibitem[{\citenamefont{Mart\'{\i}nez-Rat\'on and
  Cuesta}(1998)}]{martinez-raton:1998}
\bibinfo{author}{\bibfnamefont{Y.}~\bibnamefont{Mart\'{\i}nez-Rat\'on}}
  \bibnamefont{and} \bibinfo{author}{\bibfnamefont{J.~A.}
  \bibnamefont{Cuesta}}, \bibinfo{journal}{Phys.\ Rev.\ E}
  \textbf{\bibinfo{volume}{58}}, \bibinfo{pages}{R4080} (\bibinfo{year}{1998}).

\bibitem[{\citenamefont{Temperley}(1965)}]{temperley:1965}
\bibinfo{author}{\bibfnamefont{H.~N.~V.} \bibnamefont{Temperley}},
  \bibinfo{journal}{Proc. Phys. Soc.} \textbf{\bibinfo{volume}{86}},
  \bibinfo{pages}{185} (\bibinfo{year}{1965}).

\bibitem[{\citenamefont{Runnels}(1972)}]{runnels:1972}
\bibinfo{author}{\bibfnamefont{L.~K.} \bibnamefont{Runnels}}, in
  \emph{\bibinfo{booktitle}{Phase Transitions and Critical Phenomena}}, edited
  by \bibinfo{editor}{\bibfnamefont{C.}~\bibnamefont{Domb}} \bibnamefont{and}
  \bibinfo{editor}{\bibfnamefont{M.~S.} \bibnamefont{Green}}
  (\bibinfo{publisher}{Academic Press}, \bibinfo{address}{London},
  \bibinfo{year}{1972}), vol.~\bibinfo{volume}{2}, chap.~\bibinfo{chapter}{8},
  pp. \bibinfo{pages}{305--328}.

\bibitem[{\citenamefont{Burley}(1972)}]{burley:1972}
\bibinfo{author}{\bibfnamefont{D.~M.} \bibnamefont{Burley}}, in
  \emph{\bibinfo{booktitle}{Phase Transitions and Critical Phenomena}}, edited
  by \bibinfo{editor}{\bibfnamefont{C.}~\bibnamefont{Domb}} \bibnamefont{and}
  \bibinfo{editor}{\bibfnamefont{M.~S.} \bibnamefont{Green}}
  (\bibinfo{publisher}{Academic Press}, \bibinfo{address}{London},
  \bibinfo{year}{1972}), vol.~\bibinfo{volume}{2}, chap.~\bibinfo{chapter}{9},
  pp. \bibinfo{pages}{329--374}.

\bibitem[{\citenamefont{Evans}(1992)}]{evans:1992}
\bibinfo{author}{\bibfnamefont{R.}~\bibnamefont{Evans}}, in
  \emph{\bibinfo{booktitle}{Fundamentals of Inhomogeneous Fluids}}, edited by
  \bibinfo{editor}{\bibfnamefont{D.}~\bibnamefont{Henderson}}
  (\bibinfo{publisher}{Marcel Dekker}, \bibinfo{address}{New York},
  \bibinfo{year}{1992}), chap.~\bibinfo{chapter}{3}, p.~\bibinfo{pages}{85}.

\bibitem[{\citenamefont{Denton and Ashcroft}(1990)}]{denton:1990}
\bibinfo{author}{\bibfnamefont{A.~R.} \bibnamefont{Denton}} \bibnamefont{and}
  \bibinfo{author}{\bibfnamefont{N.~W.} \bibnamefont{Ashcroft}},
  \bibinfo{journal}{Phys. Rev. A} \textbf{\bibinfo{volume}{42}},
  \bibinfo{pages}{7312} (\bibinfo{year}{1990}).

\bibitem[{\citenamefont{Dijkstra}(1998)}]{dijkstra:1998b}
\bibinfo{author}{\bibfnamefont{M.}~\bibnamefont{Dijkstra}},
  \bibinfo{journal}{Phys.\ Rev.\ E} \textbf{\bibinfo{volume}{58}},
  \bibinfo{pages}{7523} (\bibinfo{year}{1998}).

\bibitem[{\citenamefont{Louis et~al.}(2000)\citenamefont{Louis, Finken, and
  Hansen}}]{louis:2000}
\bibinfo{author}{\bibfnamefont{A.~A.} \bibnamefont{Louis}},
  \bibinfo{author}{\bibfnamefont{R.}~\bibnamefont{Finken}}, \bibnamefont{and}
  \bibinfo{author}{\bibfnamefont{J.~P.} \bibnamefont{Hansen}},
  \bibinfo{journal}{Phys.\ Rev.\ E} \textbf{\bibinfo{volume}{61}},
  \bibinfo{pages}{R1028} (\bibinfo{year}{2000}).

\bibitem[{\citenamefont{Germain and Amokrane}(2002)}]{germain:2002}
\bibinfo{author}{\bibfnamefont{P.}~\bibnamefont{Germain}} \bibnamefont{and}
  \bibinfo{author}{\bibfnamefont{S.}~\bibnamefont{Amokrane}},
  \bibinfo{journal}{Phys. Rev. E} \textbf{\bibinfo{volume}{65}},
  \bibinfo{pages}{031109} (\bibinfo{year}{2002}).

\bibitem[{\citenamefont{Dijkstra and Frenkel}(1994)}]{dijkstra:1994}
\bibinfo{author}{\bibfnamefont{M.}~\bibnamefont{Dijkstra}} \bibnamefont{and}
  \bibinfo{author}{\bibfnamefont{D.}~\bibnamefont{Frenkel}},
  \bibinfo{journal}{Phys.\ Rev.\ Lett.} \textbf{\bibinfo{volume}{72}},
  \bibinfo{pages}{298} (\bibinfo{year}{1994}).

\bibitem[{\citenamefont{Lafuente and
  Cuesta}(2002{\natexlab{a}})}]{lafuente:2002b}
\bibinfo{author}{\bibfnamefont{L.}~\bibnamefont{Lafuente}} \bibnamefont{and}
  \bibinfo{author}{\bibfnamefont{J.~A.} \bibnamefont{Cuesta}},
  \bibinfo{journal}{J.\ Phys.: Condens.\ Matter} \textbf{\bibinfo{volume}{14}},
  \bibinfo{pages}{12079} (\bibinfo{year}{2002}{\natexlab{a}}).

\bibitem[{\citenamefont{Lafuente and
  Cuesta}(2002{\natexlab{b}})}]{lafuente:2002a}
\bibinfo{author}{\bibfnamefont{L.}~\bibnamefont{Lafuente}} \bibnamefont{and}
  \bibinfo{author}{\bibfnamefont{J.~A.} \bibnamefont{Cuesta}},
  \bibinfo{journal}{Phys.\ Rev.\ Lett.} \textbf{\bibinfo{volume}{89}},
  \bibinfo{pages}{145701} (\bibinfo{year}{2002}{\natexlab{b}}).

\bibitem[{\citenamefont{Nieswand
  et~al.}(1993{\natexlab{a}})\citenamefont{Nieswand, Dieterich, and
  Majhofer}}]{nieswand:1993a}
\bibinfo{author}{\bibfnamefont{M.}~\bibnamefont{Nieswand}},
  \bibinfo{author}{\bibfnamefont{W.}~\bibnamefont{Dieterich}},
  \bibnamefont{and} \bibinfo{author}{\bibfnamefont{A.}~\bibnamefont{Majhofer}},
  \bibinfo{journal}{Phys. Rev. E} \textbf{\bibinfo{volume}{47}},
  \bibinfo{pages}{718} (\bibinfo{year}{1993}{\natexlab{a}}).

\bibitem[{\citenamefont{Nieswand
  et~al.}(1993{\natexlab{b}})\citenamefont{Nieswand, Majhofer, and
  Dieterich}}]{nieswand:1993b}
\bibinfo{author}{\bibfnamefont{M.}~\bibnamefont{Nieswand}},
  \bibinfo{author}{\bibfnamefont{A.}~\bibnamefont{Majhofer}}, \bibnamefont{and}
  \bibinfo{author}{\bibfnamefont{W.}~\bibnamefont{Dieterich}},
  \bibinfo{journal}{Phys. Rev. E} \textbf{\bibinfo{volume}{48}},
  \bibinfo{pages}{2521} (\bibinfo{year}{1993}{\natexlab{b}}).

\bibitem[{\citenamefont{Reinel et~al.}(1994)\citenamefont{Reinel, Dieterich,
  and Majhofer}}]{reinel:1994}
\bibinfo{author}{\bibfnamefont{D.}~\bibnamefont{Reinel}},
  \bibinfo{author}{\bibfnamefont{W.}~\bibnamefont{Dieterich}},
  \bibnamefont{and} \bibinfo{author}{\bibfnamefont{A.}~\bibnamefont{Majhofer}},
  \bibinfo{journal}{Phys. Rev. E} \textbf{\bibinfo{volume}{50}},
  \bibinfo{pages}{4744} (\bibinfo{year}{1994}).

\bibitem[{\citenamefont{Tarazona and Rosenfeld}(1997)}]{tarazona:1997}
\bibinfo{author}{\bibfnamefont{P.}~\bibnamefont{Tarazona}} \bibnamefont{and}
  \bibinfo{author}{\bibfnamefont{Y.}~\bibnamefont{Rosenfeld}},
  \bibinfo{journal}{Phys.\ Rev.\ E} \textbf{\bibinfo{volume}{55}},
  \bibinfo{pages}{R4873} (\bibinfo{year}{1997}).

\bibitem[{\citenamefont{Cuesta and Mart\'{\i}nez-Rat\'on}(1997)}]{cuesta:1997a}
\bibinfo{author}{\bibfnamefont{J.~A.} \bibnamefont{Cuesta}} \bibnamefont{and}
  \bibinfo{author}{\bibfnamefont{Y.}~\bibnamefont{Mart\'{\i}nez-Rat\'on}},
  \bibinfo{journal}{Phys.\ Rev.\ Lett.} \textbf{\bibinfo{volume}{78}},
  \bibinfo{pages}{3681} (\bibinfo{year}{1997}).

\bibitem[{\citenamefont{Ree and Chesnut}(1967)}]{ree:1967}
\bibinfo{author}{\bibfnamefont{F.~H.} \bibnamefont{Ree}} \bibnamefont{and}
  \bibinfo{author}{\bibfnamefont{D.~A.} \bibnamefont{Chesnut}},
  \bibinfo{journal}{Phys. Rev. Lett.} \textbf{\bibinfo{volume}{18}},
  \bibinfo{pages}{5} (\bibinfo{year}{1967}).

\bibitem[{\citenamefont{Nisbet and Farquhar}(1974)}]{nisbet:1974}
\bibinfo{author}{\bibfnamefont{R.~M.} \bibnamefont{Nisbet}} \bibnamefont{and}
  \bibinfo{author}{\bibfnamefont{I.~E.} \bibnamefont{Farquhar}},
  \bibinfo{journal}{Physica} \textbf{\bibinfo{volume}{73}},
  \bibinfo{pages}{351} (\bibinfo{year}{1974}).

\bibitem[{\citenamefont{Bellemans and Nigam}(1967)}]{bellemans:1967}
\bibinfo{author}{\bibfnamefont{A.}~\bibnamefont{Bellemans}} \bibnamefont{and}
  \bibinfo{author}{\bibfnamefont{R.~K.} \bibnamefont{Nigam}},
  \bibinfo{journal}{J.\ Chem.\ Phys.} \textbf{\bibinfo{volume}{46}},
  \bibinfo{pages}{2922} (\bibinfo{year}{1967}).

\bibitem[{\citenamefont{Baram}(1983)}]{baram:1983}
\bibinfo{author}{\bibfnamefont{A.}~\bibnamefont{Baram}}, \bibinfo{journal}{J.
  Phys. A: Math. Gen.} \textbf{\bibinfo{volume}{16}}, \bibinfo{pages}{L19}
  (\bibinfo{year}{1983}).

\bibitem[{\citenamefont{Baram and Luban}(1987)}]{baram:1987}
\bibinfo{author}{\bibfnamefont{A.}~\bibnamefont{Baram}} \bibnamefont{and}
  \bibinfo{author}{\bibfnamefont{M.}~\bibnamefont{Luban}},
  \bibinfo{journal}{Phys. Rev. A} \textbf{\bibinfo{volume}{36}},
  \bibinfo{pages}{760} (\bibinfo{year}{1987}).

\bibitem[{\citenamefont{Slotte}(1983)}]{slotte:1983}
\bibinfo{author}{\bibfnamefont{P.~A.} \bibnamefont{Slotte}},
  \bibinfo{journal}{J. Phys. C: Solid State Phys.}
  \textbf{\bibinfo{volume}{16}}, \bibinfo{pages}{2935} (\bibinfo{year}{1983}).

\bibitem[{\citenamefont{Groh and Mulder}(2001)}]{groh:2001}
\bibinfo{author}{\bibfnamefont{B.}~\bibnamefont{Groh}} \bibnamefont{and}
  \bibinfo{author}{\bibfnamefont{B.}~\bibnamefont{Mulder}},
  \bibinfo{journal}{J.\ Chem.\ Phys.} \textbf{\bibinfo{volume}{114}},
  \bibinfo{pages}{3653} (\bibinfo{year}{2001}).

\bibitem[{\citenamefont{Rushbrooke and Scoins}(1955)}]{rushbrooke:1955}
\bibinfo{author}{\bibfnamefont{G.~S.} \bibnamefont{Rushbrooke}}
  \bibnamefont{and} \bibinfo{author}{\bibfnamefont{H.~I.}
  \bibnamefont{Scoins}}, \bibinfo{journal}{Proc. Roy. Soc.}
  \textbf{\bibinfo{volume}{A230}}, \bibinfo{pages}{74} (\bibinfo{year}{1955}).

\bibitem[{\citenamefont{Cuesta}(1999)}]{cuesta:1999}
\bibinfo{author}{\bibfnamefont{J.~A.} \bibnamefont{Cuesta}},
  \bibinfo{journal}{Europhys.\ Lett.} \textbf{\bibinfo{volume}{46}},
  \bibinfo{pages}{197} (\bibinfo{year}{1999}).

\end{thebibliography}

\begin{widetext}
\appendix*
\section{}

Let us consider a system of parallel hard cubes in
a simple cubic lattice with edge-length $\sigma=2$.
This system is equivalent to the lattice gas on a simple cubic lattice
with first, second and third neighbor exclusion. The excess
functional, in diagrammatic notation, can be written as
\begin{equation}\label{f3d}
\bFex[\rho]=\sum_{\bs\in\ZZZ}\left[\Phi_0\left(\dncb\right)-
\Phi_0\left(\dnsq\right)-\Phi_0\left(\dnxy\right)-\Phi_0\left(\dnxz\right)
+\Phi_0\left(\dnv\right)+\Phi_0\left(\dnh\right)+\Phi_0\left(\dnx\right)
-\Phi_0\left(\dnp\right)\right],
\end{equation}
where the diagrams represent the weighted densities
\begin{align*}
\dncb=n^{(1,1,1)}(\bs)&=\sum_{i,j,k=0,1}\rho(s_1+i,s_2+j,s_3+k), \quad &
\dnv=n^{(0,0,1)}(\bs)&=\sum_{i=0,1}\rho(s_1,s_2,s_3+i),\\
\dnsq=n^{(1,0,1)}(\bs)&=\sum_{i,j=0,1}\rho(s_1,s_2+i,s_3+j), &
\dnh=n^{(0,1,0)}(\bs)&=\sum_{i=0,1}\rho(s_1,s_2+i,s_3),\\
\dnxy=n^{(1,1,0)}(\bs)&=\sum_{i,j=0,1}\rho(s_1+i,s_2+j,s_3), &
\dnx=n^{(1,0,0)}(\bs)&=\sum_{i=0,1}\rho(s_1+i,s_2,s_3),\\
\dnxz=n^{(1,0,1)}(\bs)&=\sum_{i,j=0,1}\rho(s_1+i,s_2,s_3+j), &
\dnp=n^{(0,0,0)}(\bs)&=\rho(s_1,s_2,s_3).
\end{align*}

In Ref.~\onlinecite{lafuente:2002b} it was pointed out that the family
of approximate functionals constructed after the prescription of the lattice
fundamental measure theory consistently satisfied the dimensional
reduction property of the exact functionals. Hence, the functional
(\ref{f2d}) must be recovered from the functional (\ref{f3d}). To show it
we can apply an infinite external potential in every site on the tridimensional
lattice, except in the sites laying on the plane 
$\mathcal{P}=\{(0,s_2,s_3):s_2,s_3\in\Z\}$. 
Then the different contributions to the effective excess functional
become
\begin{align*}
\sum_{s_1\in\Z}\Phi_0\left(\dncb\right)&=2\Phi_0\left(\dnsq\right), \quad &
\sum_{s_1\in\Z}\Phi_0\left(\dnv\right)&=\Phi_0\left(\dnv\right),\\
\sum_{s_1\in\Z}\Phi_0\left(\dnsq\right)&=\Phi_0\left(\dnsq\right), &
\sum_{s_1\in\Z}\Phi_0\left(\dnh\right)&=\Phi_0\left(\dnh\right),\\
\sum_{s_1\in\Z}\Phi_0\left(\dnxy\right)&=2\Phi_0\left(\dnh\right), &
\sum_{s_1\in\Z}\Phi_0\left(\dnx\right)&=2\Phi_0\left(\dnp\right),\\
\sum_{s_1\in\Z}\Phi_0\left(\dnxz\right)&=2\Phi_0\left(\dnv\right), &
\sum_{s_1\in\Z}\Phi_0\left(\dnp\right)&=\Phi_0\left(\dnp\right),
\end{align*}
and therefore (\ref{f3d}) reduces to (\ref{f2d}). 

This diagrammatic notation can also be extended to multicomponent
systems, although for more than two components it becomes too
cumbersome.
\end{widetext}

\end{document}